\documentclass[english,hyper]{JHEP3}
\usepackage[T1]{fontenc}
\usepackage[latin9]{inputenc}
\usepackage{booktabs}
\usepackage{units}
\usepackage{amsmath}
\usepackage{graphicx}
\usepackage{amssymb}

\makeatletter

\DeclareRobustCommand*{\lyxarrow}{%
\@ifstar
{\leavevmode\,$\triangleleft$\,\allowbreak}
{\leavevmode\,$\triangleright$\,\allowbreak}}
\providecommand{\tabularnewline}{\\}

\usepackage{cite}

\title{Neutrino masses from new generations}
\author{Alberto Aparici\footnote{alberto.aparici@uv.es}, 
Juan Herrero-García\footnote{juan.a.herrero@uv.es},
Nuria Rius\footnote{nuria@ific.uv.es}\,
and Arcadi Santamaria\footnote{arcadi.santamaria@uv.es}  
\\
Depto.\ de Física Teòrica,
and IFIC, Universidad de
Valencia-CSIC \\ 
Edificio de Institutos de Paterna, Apt. 22085, 46071 Valencia,
Spain}

\abstract{We reconsider the possibility that Majorana masses for the three 
known neutrinos are generated radiatively by the presence of a fourth 
generation and one right-handed neutrino  with Yukawa 
couplings and a Majorana mass term.
We find that the observed light neutrino mass hierarchy is not compatible 
with low energy universality bounds in this minimal 
scenario, but all present data can be accommodated with five generations
and two right-handed neutrinos.  
Within this framework, we explore the parameter space regions which 
are currently allowed and could lead to observable effects in 
neutrinoless double beta decay, $\mu - e$ conversion in nuclei and  
$\mu \rightarrow e \gamma$ experiments. We also discuss the detection prospects at LHC.}  

\preprint{IFIC/11-11\\
FTUV-11-0414}

\makeatother

\usepackage{babel}

\begin{document}

\section{Introduction}

Neutrino oscillation data \cite{Cleveland:1998nv,Hampel:1998xg,Altmann:2000ft,Ahmad:2002jz,Eguchi:2002dm,Fukuda:1998mi,Allison:1999ms,Aliu:2004sq,Davis:1968cp,Hirata:1988uy,Bellini:2008mr,Adamson:2009yc,Schwetz:2008er,GonzalezGarcia:2007ib}
require at least two massive neutrinos with large mixing, providing
one of the strongest evidences of physics beyond the Standard Model
(SM). However, the new physics scale responsible for neutrino masses
is largely unknown. With the starting of the LHC, new physics scales
of order TeV will become testable through direct production of new
particles, so it is very interesting to explore low-energy scenarios
for neutrino masses. Moreover, typically, these scenarios also lead
to observable signatures in precision experiments, such as violations
of universality, charged lepton flavour violating (LFV) rare decays
such as $\ell_{i}\rightarrow\ell_{j}\gamma$ or $\mu-e$ conversion in
nuclei, which, being complementary to the LHC measurements, may help
to discriminate among different models. Regarding the fundamental
question of the neutrino mass nature, Dirac or Majorana, 
lepton-number-violating
low-scale models may give additional contributions to neutrinoless
double beta ($0\nu2\beta$) decay process, shedding new light on this
issue.

On the other hand, one of the most natural extensions of the SM that
has been extensively explored in the last years is the addition of
one (or more) sequential generations of quarks and leptons
\cite{Frampton:1999xi}. This extension
is very natural and has a rich phenomenology both at LHC as well as
in LFV processes. Moreover, new generations address some of the open
questions in the SM and can accommodate emerging hints on new physics
(see for instance \cite{Holdom:2009rf} for a recent review). 

Theoretically, apart from simplicity, there are no compelling arguments
in favour of only three families. In theories with extra dimensions
one can relate the number of families to the topology of the compact
extra dimensions or set constraints on the number of chiral families
and allowed gauge groups by requiring anomaly cancellation. Then,
one can build models to justify only three generations at low energies.
However, one could also build other models in order to justify four
or more generations. In the SM in four dimensions anomalies cancel
within each generation and, therefore, the number of families is in
principle free. 

From the phenomenological point of view it seems that the most striking
argument against new generations is the measurement of the invisible
$Z$-boson decay width, $\Gamma_{\mathrm{inv}}$, which effectively
counts the number of light degrees of freedom coupled to the $Z$-boson
(lighter than $m_{Z}/2$) which is very close to $3$ \cite{Nakamura:2010zzi}.
However, if neutrinos from new families are heavy they do not contribute
to $\Gamma_{\mathrm{inv}}$ and, then, additional generations are
allowed. Still, pairs of virtual heavy fermions from new generations
contribute to the electroweak parameters and spoil the agreement of
the SM with experiment. Global fits of models with additional generations
to the electroweak data have been performed 
\cite{He:2001tp,Novikov:2009kc}
and the conclusion is that they favour no more that five generations
with appropriate masses for the new particles. Although some controversy
exists on the interpretation of the data (see for instance \cite{Erler:2010sk})
most of the fits make some simplifying assumptions on the mass spectrum
of the new generations and do not consider Majorana neutrino masses
for the new generations or the possibility of breaking dynamically
the gauge symmetry via the condensation of the new generations' fermions;
all these will give additional contributions to the oblique parameters
and will modify the fits. Therefore, in view that soon we will see
or exclude new generations thanks to the LHC, it is wise to approach
this possibility with an open mind.

From the discussion above, it seems that neutrinos from new generations
are very different from the ones discovered up to now, since they
should have a mass $10^{11}$ times larger. However, this apparent
difference is naturally explained within the framework that we are
going to explore. In the SM neutrinos are massless because there are
no right-handed neutrinos and because, with the minimal Higgs sector,
lepton number is automatically conserved. We now know that neutrinos
have masses, therefore the SM has to be modified to accommodate them;
the simplest possibility is to add three right-handed neutrinos with
Dirac mass terms, like for the rest of the fermions in the SM. If
one then considers the SM with four generations (and four right-handed
neutrinos), it is very difficult to justify why the neutrino from
the fourth generation is $10^{11}$ times heavier than the three observed
ones. This difficulty is alleviated if right-handed neutrinos have
Majorana masses at the electroweak scale and the Dirac masses of the
neutrinos are of the order of magnitude of their corresponding charged
leptons \cite{Hill:1989vn}. Then, the see-saw mechanism is operative
and gives neutrino masses $m_{1}\sim m_{e}^{2}/M$, $m_{2}\sim m_{\mu}^{2}/M$,
$m_{3}\sim m_{\tau}^{2}/M$, $m_{4}\sim m_{E}^{2}/M\sim m_{E}$ (we
denote by $E$ the fourth generation charged lepton). Although with
a common Majorana mass $M$ at the electroweak scale it is not possible
to obtain $m_{3}$ light enough to fit the observed neutrino masses,
this could be solved by allowing different Majorana masses for the
different generations; but then one should explain why $M_{2},M_{3}\gg M_{4}$.

Right-handed neutrinos, however, do not have gauge charges and are
not needed to cancel anomalies, therefore their number is not linked
to the number of generations. In fact, an extension of the SM with
four generations and just one right-handed neutrino with both Dirac
and a Majorana masses at the electroweak scale leads, at tree level,
to three massless and two heavy Majorana neutrinos. Since lepton number
is broken in the model, the three massless neutrinos acquire Majorana
masses at two loops therefore providing a natural explanation for
the tiny masses of the three known neutrinos \cite{Babu:1988ig} %
\footnote{Two-loop quantum corrections within the SM with only two massive Majorana
neutrinos also lead to a (tiny) mass for the third one \cite{Davidson:2006tg}.%
}. More generally, it has been shown that in the SM with $n_{L}$ lepton
doublets, $n_{H}$ Higgs doublets and $n_{R}<n_{L}$ right-handed
neutrino singlets with Yukawa and Majorana mass terms there are $n_{L}-n_{R}$
massless Majorana neutrinos at tree level, of which $n_{L}-n_{R}-{\rm max}(0,n_{L}-n_{H}n_{R})$
states acquire mass by neutral Higgs exchange at one loop \cite{Grimus:1989pu,Branco:1988ex,Petcov:1984nz}.
The remaining ${\rm max}(0,n_{L}-n_{H}n_{R})$ states get masses at
two loops. Similar extensions could be built with additional hyperchargeless
fermion triplets, like in type III see-saw.

In this work we reconsider the model of ref. \cite{Babu:1988ig},
without enlarging the scalar sector of the SM but allowing for extra
generations. The paper is organised as follows. In section~\ref{sec:frame}
we summarize current neutrino data and searches for new generations.
In section~\ref{SM4} we review the radiative neutrino mass generation
at two loops, and show that the observed light neutrino mass hierarchy
can not be accommodated in the minimal scenario with four generations.
In section~\ref{SM5} we present a five generation example which
leads to the observed neutrino masses and (close to tribimaximal)
mixing. We introduce a simple parametrization of the model and explore
the parameter space allowed by current neutrino data, universality,
charged lepton flavour violating rare decays $\ell_{i}\rightarrow\ell_{j}\gamma$
and $0\nu2\beta$ decay, as well as the regions that will be probed
in near future experiments (MEG, $\mu-e$ conversion in nuclei). Section~\ref{colliders}
is devoted to collider phenomenology and we summarize our results
in section~\ref{conclusions}.

\section{Framework and review\label{sec:frame}}

It has been well established in the last decade that neutrinos are
massive, thanks to the results obtained with solar \cite{Cleveland:1998nv,Hampel:1998xg,Altmann:2000ft,Ahmad:2002jz,Bellini:2008mr},
and atmospheric \cite{Hirata:1988uy,Fukuda:1998mi,Allison:1999ms}
neutrinos, confirmed in experiments using man-made beams: neutrinos
from nuclear reactors \cite{Eguchi:2002dm} and accelerators \cite{Aliu:2004sq,Adamson:2009yc}.

The minimum description of all neutrino data requires mixing among
the three neutrino states with definite flavour ($\nu_{e},\nu_{\mu},\nu_{\tau}$),
which can be expressed as quantum superpositions of three massive
states $\nu_{i}$ (i=1,2,3) with masses $m_{i}$. The standard parametrization
of the leptonic mixing matrix, $U_{{\rm PMNS}}$, is: \begin{eqnarray}
U_{{\rm PMNS}}=\left(\begin{array}{ccc}
c_{13}c_{12} & c_{13}s_{12} & s_{13}e^{-i\delta}\\
-c_{23}s_{12}-s_{23}s_{13}c_{12}e^{i\delta} & c_{23}c_{12}-s_{23}s_{13}s_{12}e^{i\delta} & s_{23}c_{13}\\
s_{23}s_{12}-c_{23}s_{13}c_{12}e^{i\delta} & -s_{23}c_{12}-c_{23}s_{13}s_{12}e^{i\delta} & c_{23}c_{13}\end{array}\right)\left(\begin{array}{ccc}
e^{i\phi_{1}}\\
 & e^{i\phi_{2}}\\
 &  & 1\end{array}\right)\ ,\label{UPMNS}\end{eqnarray}
where $c_{ij}\equiv\cos\theta_{ij}$ and $s_{ij}\equiv\sin\theta_{ij}$.
In addition to the Dirac-type phase $\delta$, analogous to that of
the quark sector, there are two physical phases $\phi_{i}$ if neutrinos
are Majorana particles. The measurement of these parameters is by
now restricted to oscillation experiments which are only sensitive
to mass-squared splittings ($\Delta m_{ij}^{2}\equiv m_{i}^{2}-m_{j}^{2}$).
Moreover, oscillations in vacuum cannot determine the sign of the
splittings. As a consequence, an uncertainty in the ordering of the
masses remains; the two possibilities are: 

\begin{eqnarray}
m_{1}<m_{2}<m_{3},\\
m_{3}<m_{1}<m_{2}.\end{eqnarray}
 The first option is the so-called normal hierarchy spectrum while
the second one is the inverted hierarchy scheme; in this form they
correspond to the two possible choices of the sign of $\Delta m_{31}^{2}\equiv\Delta m_{atm}^{2}$,
which is still undetermined, while $\Delta m_{21}^{2}\equiv\Delta m_{sol}^{2}$
is known to be positive. Within this minimal context, two mixing angles
and two mass-squared splittings are relatively well determined from
oscillation experiments (see table \ref{tab:table1}), there is a
slight hint of $\theta_{13}>0$ and nothing is known about the phases.

\TABLE{
\centering{}\begin{tabular}{c}
\toprule 
Light neutrino best fit values\tabularnewline
\midrule 
\addlinespace[6pt]
$\Delta m_{21}^{2}=(7.64{+0.19\atop -0.18})\times10^{-5}\:\textrm{eV}^{2}$\tabularnewline\addlinespace[6pt]
\addlinespace[6pt]
$\Delta m_{31}^{2}=\,\,\,\,\begin{cases}
\:(2.45\pm0.09)\times10^{-3}\:\textrm{eV}^{2} & \mathrm{NH}\\
-(2.34{+0.10\atop -0.09})\times10^{-3}\:\textrm{eV}^{2} & \mathrm{IH}\end{cases}$ \tabularnewline\addlinespace[6pt]
\addlinespace[6pt]
$\sin^{2}\theta_{12}=0.316\pm0.016$ \tabularnewline\addlinespace[6pt]
\addlinespace[6pt]
$\sin^{2}\theta_{23}=\,\begin{cases}
\,0.51\pm0.06 & \mathrm{NH}\\
\,0.52\pm0.06 & \mathrm{IH}\end{cases}$ \tabularnewline\addlinespace[6pt]
\addlinespace[6pt]
$\sin^{2}\theta_{13}=\,\begin{cases}
\,0.017{+0.007\atop -0.009} & \mathrm{NH}\\
\,0.020{+0.008\atop -0.009} & \mathrm{IH}\end{cases}$ \tabularnewline\addlinespace[6pt]
\bottomrule
\end{tabular}\caption{The best fit values of the light neutrino parameters and their $1\sigma$
errors from \cite{Schwetz:2011qt}. \label{tab:table1} }
}

Regarding the absolute neutrino mass scale, it is constrained by laboratory
experiments searching for its kinematic effects in Tritium $\beta$-decay,
which are sensitive to the so-called effective electron neutrino mass,
\begin{equation}
m_{\nu_{e}}^{2}\equiv\sum_{i}m_{i}^{2}|U_{ei}|^{2}.\label{eq:tritium-mass}\end{equation}
 The present upper limit is $m_{\nu_{e}}<2.2$ eV at 95\% confidence
level (CL) \cite{Bonn:2001tw,Lobashev:2001uu}, while a new experimental
project, KATRIN \cite{Osipowicz:2001sq}, is underway, with an estimated
sensitivity limit $m_{\nu_{e}}\sim0.2$ eV. However, cosmological
observations provide the tightest constraints on the absolute scale
of neutrino masses, via their contribution to the energy density of
the Universe and the growth of structure. In general these bounds
depend on the assumptions made about the expansion history as well
as on the cosmological data included in the analysis \cite{Lesgourgues:2006nd}.
Combining CMB and large scale structure data quite robust bounds have
been obtained: $\sum_{i}m_{i}$ < 0.4 eV at 95\% CL within the $\Lambda$CDM
model \cite{Reid:2009nq} and $\sum_{i}m_{i}<1.5\,\mathrm{eV}$ at
95\%~CL when allowing for several departures from $\Lambda$CDM \cite{GonzalezGarcia:2010un}.

Finally, if neutrinos are Majorana particles complementary information
on neutrino masses can be obtained from 0$\nu2\beta$ decay. The contribution
of the known light neutrinos to the $0\nu2\beta$ decay amplitude
is proportional to the effective Majorana mass of $\nu_{e}$, $m_{ee}=|\sum_{i}m_{i}U_{ei}^{2}|$,
which depends not only on the masses and mixing angles of the $U_{PMNS}$
matrix but also on the phases. The present bound from the Heidelberg-Moscow
group is $m_{ee}<0.34$ eV at 90\% CL \cite{KlapdorKleingrothaus:2000sn},
but future experiments can reach sensitivities of up to $m_{ee}\sim0.01$
eV \cite{Avignone:2007fu}.

We now briefly review the current status of searches for new sequential
generations. Direct production of the 4$^{\mathrm{th}}$ generation
quarks $t'$ and $b'$, assuming $t'\rightarrow Wq$ and $b'\rightarrow Wt$
has been searched in CDF, leading to the lower mass bounds $m_{t'}>335$
GeV and $m_{b'}>385$ GeV \cite{Lister:2008is,Flacco:2011ym}. Limits
on new generation leptons, from LEP II, are weaker: $m_{\ell'}>100.8$
GeV and $m_{\nu'}>$ $80.5\:(90.3)\:\mathrm{GeV}$ for pure Majorana
(Dirac) particles, assuming that the 4$^{\mathrm{th}}$ generation
leptons are unstable, i.e., their mixing with the known leptons is
large enough so that they decay inside the detector \cite{Nakamura:2010zzi}.
When neutrinos have both Dirac and Majorana masses, their coupling
to the $Z$ boson may be reduced by the neutrino mixing angle and
the bound on the lightest neutrino mass may be relaxed to $63\,\mathrm{GeV}$
\cite{Carpenter:2010dt}. While the bound on a charged lepton stable
on collider lifetimes is still about $100$~GeV, in the case of stable
neutrinos the only limit comes from the LEP I measurement of the invisible
$Z$ width, $m_{\nu'}>39.5\,(45)\,\mathrm{GeV}$ for pure Majorana
(Dirac) particles \cite{Nakamura:2010zzi}.

Even if new generation fermions are very heavy and cannot de directly
produced, they affect electroweak observables through radiative corrections.
Recent works have shown that a fourth generation is consistent with
electroweak precision observables \cite{Kribs:2007nz,Erler:2010sk,Eberhardt:2010bm},
provided there is a heavy Higgs and the mass splittings of the new
SU(2) doublets satisfy \cite{Eberhardt:2010bm} %
\footnote{The allowed quark mass splittings depend on the Higgs mass, according
to the approximate formula $m_{t'}-m_{b'}\simeq\left(1+\frac{1}{5}\log(\frac{m_{H}}{115~{\rm GeV}})\right)\times50~{\rm GeV}$
from ref. \cite{Kribs:2007nz}.%
} \begin{eqnarray}
|m_{t'}-m_{b'}| & < & 80\,\,{\rm GeV},\\
|m_{\ell'}-m_{\nu'}| & < & 140\,\,{\rm GeV}\ .\end{eqnarray}
 Notice, however, that a long-lived fourth generation can reopen a
large portion of the parameter space \cite{Murayama:2010xb}.

In addition to these phenomenological bounds one can place some upper
limits by using perturbative unitarity, triviality and by imposing
the stability of the Higgs potential at one loop. Typically one obtains
limits of the order of the TeV~\cite{Chanowitz:1978mv} for degenerate
lepton doublets and about $600\,\mathrm{GeV}$ for degenerate quark
doublets. 

A very striking effect of new generations is the enhancement of the
Higgs-gluon-gluon vertex which arises from a triangle diagram with
all quarks running in the loop. This vertex is enhanced approximately
by a factor 3 (5) in the presence of a heavy fourth (fifth) generation
\cite{Kribs:2007nz,Anastasiou:2010bt}. Therefore, the Higgs production
cross section through gluon fusion at the Tevatron and the LHC is
enhanced by a factor of 9 (25) in the presence of a fourth (fifth)
generation. Thus, a combined analysis from CDF and D0 for four generations
has excluded a SM-like Higgs boson with mass between $131\,\mathrm{GeV}$
and $204\,\mathrm{GeV}$ at 95\% CL \cite{Aaltonen:2010sv}, while
LHC data already excludes $144\,\mathrm{GeV}<m_{H}<207\,\mathrm{GeV}$
at 95\% CL \cite{Chatrchyan:2011tz}. From these results, we estimate
roughly that $m_{H}>300\,\mathrm{GeV}$ in the case of five generations.
However, these limits may be softened if the fourth generation neutrinos
are long-lived and the branching ratio of the decay channel $H\rightarrow\nu_{4}\bar{\nu}_{4}$
is significant \cite{Keung:2011zc}.

Putting all together, a general analysis seems to suggest that at
most only two extra generations are allowed \cite{Novikov:2009kc}
unless new additional physics is invoked. If extra generations exist,
the Higgs should be heavy. Extra generation quarks should also be
quite heavy and be almost degenerate within a generation. The constraints
on new generation leptons are milder; charged lepton and Dirac neutrino
masses should be in the range $100$--$1000\,\mathrm{GeV}$ and, as
we will see in section~\ref{pheno}, this range will increase if
neutrinos have both Dirac and Majorana mass terms.

\section{Four generations\label{SM4}}

If we add one right-handed neutrino $\nu_{R}$ to the SM with three
generations and we do not impose lepton number conservation, so that
there is a Majorana mass term for the right-handed neutrino, a particular
linear combination of $\nu_{e},\,\nu_{\mu},\,\nu_{\tau}$, call it
$\nu{}_{3}^{\prime}$, will couple to $\nu_{R}$ and get a Majorana
mass at tree level. The other two linear combinations are massless
at tree level but, since lepton number is broken, no symmetry protects
them from acquiring a Majorana mass at the quantum level. In fact,
they obtain a mass at two loops by the exchange of two $W$ bosons
(same diagram as in figure~\ref{fig:TwoW}, but with $\nu_{3},\nu_{\bar{3}}$
running in the loop)\textbf{.} This leads to two extremely small
neutrino masses, as desired, but there is a huge hierarchy between
the tree-level mass, for $\nu_{3}$, and the two-loop-level masses,
for $\nu_{1}$ and $\nu_{2},$ therefore this possibility cannot accommodate
the observed neutrino masses.%
\footnote{See however \cite{Grimus:1999wm} for a model with three generations,
one right-handed neutrino singlet and two Higgs doublets which can
accommodate neutrino masses and mixings.%
}

\FIGURE{
\centering{}\includegraphics[width=0.60\columnwidth]{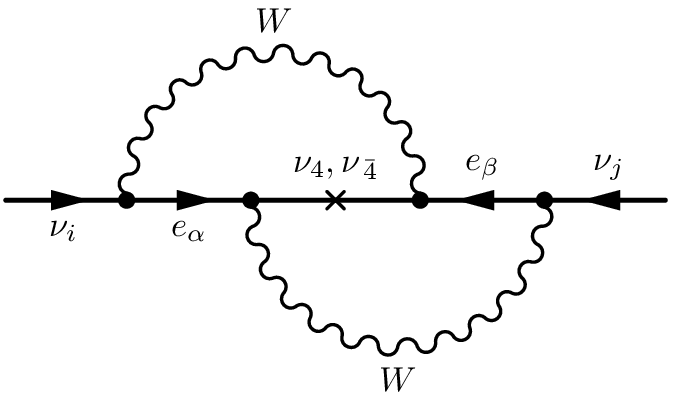}\caption{Two-loop diagram contributing to neutrino masses in the four-generation
model. \label{fig:TwoW}}
}

Analogously, we can extend the SM by adding a complete fourth generation
and one right handed neutrino $\nu_{R}$ with a Majorana mass term
\cite{Petcov:1984nz,Babu:1988ig,Babu:1988wk,Choudhury:1994vr}. We
denote the new charged lepton $E$ and the new neutrino $\nu_{E}$.
The relevant part of the Lagrangian is

\begin{equation}
\mathcal{L}_{Y}=-\bar{\ell}Y_{e}e_{R}\phi-\bar{\ell}Y_{\nu}\nu_{R}\tilde{\phi}-\cfrac{1}{2}\,\overline{\nu_{R}^{c}}m_{R}\nu_{R}+\mathrm{H.c.}\ ,\label{eq:4genlag}\end{equation}
where $\ell$ represents the left-handed lepton SU(2) doublets, $e_{R}$
the right-handed charged leptons, $\nu_{R}$ the right-handed singlet
and flavour indices are omitted. In generation space $\ell$ and $e_{R}$
are organized as column vectors with four components. Thus, $Y_{e}$
is a general, $4\times4$ matrix, $Y_{\nu}$ is a general four-component
column vector whose elements we denote by $y_{\alpha}$ with $\alpha=e,\mu,\tau,E$,
and $m_{R}$ is a Majorana mass term. The standard kinetic terms,
not shown in eq.~(\ref{eq:4genlag}), are invariant under general
unitary transformations $\ell\rightarrow V_{\ell}\ell$, $e_{R}\rightarrow V_{e}e_{R}$
and $\nu_{R}\rightarrow e^{i\alpha}\nu_{R}$. One can use those transformations,
$V_{\ell}$ and $V_{e}$, to choose $Y_{e}$ diagonal and positive
and also $m_{R}$ can be taken positive by absorbing its phase in
$\nu_{R}$. $Y_{\nu}$ is in general arbitrary; however, there is
still a rephasing invariance in $\ell$ and $e_R$ that will allow us to remove
all phases in $Y_{\nu}$.

After spontaneous symmetry breaking (SSB) the mass matrix for the neutral 
leptons is a $5\times5$
Majorana symmetric matrix which has the standard see-saw structure
with only one right-handed neutrino Majorana mass term. Therefore,
it leads to two massive Majorana and three massless Weyl neutrinos.
From the Lagrangian it is clear that only the linear combination of
left-handed neutrinos $\nu_{4}^{\prime}\propto y_{e}\nu_{e}+y_{\mu}\nu_{\mu}+y_{\tau}\nu_{\tau}+y_{E}\nu_{E}$
will pair up with $\nu_{R}$ to acquire a Dirac mass term. Thus, it
is convenient to pass from the flavour basis ($\nu_{e},\nu_{\mu},\nu_{\tau},\nu_{E}$)
to a new one $\nu_{\text{1}}^{\prime},\nu_{\text{2}}^{\prime},\nu_{\text{3}}^{\prime},\nu_{\text{4}}^{\prime}$
where the first three states will be massless at tree level and only
$\nu_{4}^{\prime}$ will mix with $\nu_{R}$. If $V$ is the orthogonal
matrix that passes from one basis to the other we will have $\nu_{\alpha}=\sum_{i}V_{\alpha i}\nu_{i}^{\prime}$
($i=1,\text{\ensuremath{\cdots}},4$, $\alpha=e,\mu,\tau,E$) with
$V_{\alpha4}\text{\ensuremath{\equiv}}N_{\alpha}=y_{\alpha}/\sqrt{\sum_{\beta}y_{\beta}^{2}}$.
Since $\nu_{\text{1}}^{\prime},\nu_{\text{2}}^{\prime},\nu_{\text{3}}^{\prime}$
are massless, we are free to choose them in any combination of $\nu_{e},\nu_{\mu},\nu_{\tau},\nu_{E}$
as long as they are orthogonal to $\nu_{4}^{\prime}$, i.e., $\sum_{\alpha}V_{\text{\ensuremath{\alpha}i}}N_{\alpha}=0$
for $i\text{=1,2,3}$. The orthogonality of $V$ almost fixes all
its elements in terms of $N_{\alpha}$, but still leaves us some freedom
to set three of them to zero. Following \cite{Babu:1988wk,Babu:1988ig}
we choose $V_{\tau1}=V_{E1}=V_{E2}=0$ for convenience.

After this change of basis, we are left with a non-trivial $2\times2$
mass matrix for $\nu_{4}^{\prime}$ and $\nu_{R}$ which can easily
be diagonalized and leads to two Majorana neutrinos\[
\nu_{4}=i\cos\theta(-\nu_{4}^{\prime}+\nu_{4}^{\prime c})+i\sin\theta(\nu_{R}-\nu_{R}^{c}),\]
 \[
\nu_{\bar{4}}=-\sin\theta(\nu_{4}^{\prime}+\nu_{4}^{\prime c})+\cos\theta(\nu_{R}+\nu_{R}^{c}),\]
 \begin{equation}
m_{4,\bar{4}}=\cfrac{1}{2}\left(\sqrt{m_{R}{}^{2}+4m_{D}^{2}}\mp m_{R}\right),\label{eq:masses-mixings-4gens}\end{equation}
 where $m_{D}=v\sqrt{\sum_{i}y_{i}^{2}}$, with $v=\langle\phi^{(0)}\rangle$,
and $\tan^{2}\theta=m_{4}/m_{\bar{4}}$. The factor $i$ and the relative
signs in $\nu_{4}$ are necessary to keep the mass terms positive
and preserve the canonical Majorana condition $\nu_{4}=\nu_{4}^{c}$.
If $m_{R}\ll m_{D}$, we have $m_{4}\approx m_{\bar{4}}$, $\tan\theta\approx1$,
and we say we are in the pseudo-Dirac limit  while when $m_{R}\gg m_{D}$,
$m_{4}\approx m_{D}^{2}/m_{R}$ and $m_{\bar{4}}\approx m_{R}$, $\tan\theta\approx m_{D}/m_{R}$
and we say we are in the see-saw limit.

Since lepton number is broken by the $\nu_{R}$ Majorana mass term,
there is no symmetry which prevents the tree-level massless neutrinos
from gaining Majorana masses at higher order. In fact, Majorana masses
for the light neutrinos, $\nu_{\text{1}}^{\prime},\nu_{\text{2}}^{\prime},\nu_{\text{3}}^{\prime}$,
are generated at two loops by the diagram of figure~\ref{fig:TwoW},
and are given by

\begin{equation}
M_{ij}=-\frac{g^{4}}{m_{W}^{4}}m_{R}m_{D}^{2}\sum_{\text{\ensuremath{\alpha}}}V_{\alpha i}V_{\alpha4}m_{\alpha}^{2}\sum_{\beta}V_{\beta j}V_{\beta4}m_{\beta}^{2}I_{\alpha\beta},\label{eq:masasligeros4gen}\end{equation}
 where the sums run over the charged leptons $\alpha,\beta=e,\mu,\tau,E$
while $i,j=1,2,3$, and

\begin{equation}
I_{\alpha\beta}=J(m_{4},m_{\bar{4}},m_{\alpha},m_{\beta},0)-\frac{3}{4}J(m_{4},m_{\bar{4}},m_{\alpha},m_{\beta},m_{W}),\label{eq:integral}\end{equation}
with $J(m_{4},m_{\bar{4}},m_{\alpha},m_{\beta},m_{W})$ the two-loop
integral defined and computed in appendix~\ref{sec:appendixA}. 

When $m_{R}=0,$ $M_{ij}=0$, as it should, because in that case lepton
number is conserved. Also when $m_{D}=0$ we obtain $M_{ij}=0$, since
then the right-handed neutrino decouples completely and lepton number
is again conserved.

To see more clearly the structure of this mass matrix we can take,
for the moment, the limit $m_{e}=m_{\mu}=m_{\tau}=0$; then, since
we have chosen $V_{\tau1}=V_{E1}=V_{E2}=0$, the only non-vanishing
element in $M_{ij}$ is $M_{33}$ and it is proportional to 
$V_{E3}^{2}N_{E}^{2}m_{E}^{4}I_{EE}$.
Keeping all the masses one can easily show that the eigenvalues of
the light neutrino mass matrix are proportional to 
$m_{\mu}^{4},\, m_{\tau}^{4},\, m_{E}^{4}$
which gives a huge hierarchy between neutrino masses. 
Moreover, for 
$m_{E}\gg m_{4,\bar{4}}\gg m_{W}$, the loop integrals in 
eq.~\eqref{eq:integral}
can be well approximated by (see appendix \ref{sec:appendixA}): 
\begin{equation}
I_{EE}\approx\cfrac{-1}{(4\pi)^{4}2m_{E}^{2}}\ln\cfrac{m_{E}}{m_{\bar{4}}}
\label{eq:intsimplificadaE4gens}\end{equation}
 and\begin{equation}
I_{\mu\mu}\approx I_{\tau\tau}\approx
\cfrac{-1}{(4\pi)^{4}2m_{\bar{4}}^{2}}\ln\cfrac{m_{\bar{4}}}{m_{4}}\ ,
\label{eq:intsimplificadatau4gens}\end{equation}
leading to only two light neutrino masses, since the mass matrix in 
eq.~(\ref{eq:masasligeros4gen}) has rank 2 
if the three light charged lepton masses are neglected in 
$I_{\alpha \beta}$. 
The third light neutrino mass is generated when at least $m_\tau$ 
is taken into account 
in the loop integral, leading to a further suppression.  
Within the above approximation, 
the following ratio of $\nu_{2}$ and $\nu_{3}$ masses is obtained
\cite{Babu:1989pz}:
\begin{equation}
\cfrac{m_{2}}{m_{3}}\lesssim\cfrac{1}{4N_{E}^{2}}\left(\cfrac{m_{\tau}}{m_{E}}\right)^{2}\left(\cfrac{m_{\tau}}{m_{\bar{4}}}\right)^{2}\lesssim\cfrac{10^{-7}}{N_{E}^{2}}\ ,\label{eq:ratiomasasligerosapprox}\end{equation}
where we have taken $\ln(m_{\bar{4}}/m_{4})\approx\ln(m_{E}/m_{\bar{4}})\approx1$
and in the last step we used that $m_{E},m_{\bar{4}}\gtrsim100\,\mathrm{GeV}$.
To overcome this huge hierarchy one would need very small values of
$N_{E}$ which would imply that the heavy neutrinos are not mainly
$\nu_{E}$ but some combination of the three known neutrinos $\nu_{e},\nu_{\mu},\nu_{\tau}$;
but this is not possible since it would yield observable effects in
a variety of processes, like $\pi\rightarrow\mu\nu$, $\pi\rightarrow e\nu$,
$\tau\rightarrow e\nu\nu$, $\tau\rightarrow\mu\nu\nu$. This requires
that $y_{e,\mu,\tau}\lesssim10^{-2}y_{E}$ \cite{Buras:2010cp,Lacker:2010zz}
and then $N_{E}\approx1$.

Therefore, although the idea is very attractive, the simplest version
is unable to accommodate the observed spectrum of neutrino masses
and mixings. 
However, notice that whenever a new generation and
a right-handed neutrino with Majorana mass at (or below) the TeV scale
are added to the SM, the two-loop contribution to neutrino masses
is always present and provides an important constraint for
this kind of SM extensions.

In the following we modify the original idea by adding
one additional generation and one additional fermion singlet. We will
see that this minimal modification is able to accommodate all current
data.

\section{Five generation working example\label{SM5}}

\subsection{The five generations model}

We add two generations to the SM and two right-handed neutrinos. We
denote the two charged leptons by $E$ and $F$ and the two right-handed
singlets by $\nu_{4R}$ and $\nu_{5R}$. The Lagrangian is exactly
the same we used for four generations \eqref{eq:4genlag} but now
$\ell$ and $e$ are organized as five-component column vectors while
$\nu_{R}$ is a two-component column vector containing $\nu_{4R}$
and $\nu_{5R}$. Thus, $Y_{e}$ is a general, $5\times5$ matrix,
$Y_{\nu}$ is a general $5\times2$ matrix and $m_{R}$ is now a general
symmetric $2\times2$ matrix. The kinetic terms are invariant under
general unitary transformations $\ell\rightarrow V_{\ell}\ell$, $e_{R}\rightarrow V_{e}e_{R}$
and $\nu\rightarrow V_{\nu}\nu$, which can be used to choose $Y_{e}$
diagonal and positive and $m_{R}$ diagonal with positive elements
$m_{4R}$ and $m_{5R}$. After this choice, there is still some rephasing
invariance $\ell_{i}\rightarrow e^{i\alpha_{i}}\ell_{i}$, $e_{iR}\rightarrow e^{i\alpha_{i}}e_{iR}$
broken only by $Y_{\nu}$, which can be used to remove five phases
in $Y_{\nu}$. Therefore \begin{equation}
Y_{\nu}=\left(\begin{array}{cc}
y, & y^{\prime}\end{array}\right),\label{eq:yukawas}\end{equation}
 where $y$ and $y^{\prime}$ are five-component column vectors with
components $y_{\alpha}$ and $y_{\alpha}^{\prime}$ respectively ($\alpha=e,\mu,\tau,E,F$),
one of which can be taken real while the other, in general, will contain
phases. The model, contrary to the four-generation case, has additional
sources of CP violation in the leptonic sector. However, since at
the moment we are not interested in CP violation, for simplicity we
will take all $y_{\alpha}$ and $y_{\alpha}^{\prime}$ real.

Much as in the four-generation case, the linear combination $\nu_{4}^{\prime}\propto\sum_{\alpha}y_{\alpha}\nu_{\alpha}$
only couples to $\nu_{4R}$ and the combination $\nu_{5}^{\prime}\propto\sum_{\alpha}y_{\alpha}^{\prime}\nu_{\alpha}$
only couples to $\nu_{5R}$. Therefore, the tree-level spectrum will
contain three massless neutrinos (the linear combinations orthogonal
to $\nu_{4}^{\prime}$ and $\nu_{5}^{\prime}$) and four heavy Majorana
neutrinos. Unfortunately, since in the general case $\nu_{4}^{\prime}$
and $\nu_{5}^{\prime}$ may not be orthogonal to each other, the diagonalization
becomes much more cumbersome than in the four-generation case. Since
we just want to provide a working example, we choose $\nu_{4}^{\prime}$
and $\nu_{5}^{\prime}$ orthogonal to each other, i.e., $\sum_{\alpha}y_{\alpha}y_{\alpha}^{\prime}=0$.
This simplifies enormously the analysis of the model and allows us
to adopt a diagonalization procedure analogous to the one followed
in the four-generation case.

We change from the flavour fields $\nu_{e},\nu_{\mu},\nu_{\tau},\nu_{E},\nu_{F}$
to a new basis $\nu_{\text{1}}^{\prime},\nu_{\text{2}}^{\prime},\nu_{\text{3}}^{\prime},\nu_{\text{4}}^{\prime},\nu_{5}^{\prime}$
where $\nu_{\text{1}}^{\prime},\nu_{\text{2}}^{\prime},\nu_{\text{3}}^{\prime}$
are massless at tree level, so we are free to choose them in any combination
of the flavour states as long as they are orthogonal to $\nu_{4}^{\prime}$
and $\nu_{5}^{\prime}$. Thus, if $V$ is the orthogonal matrix that
passes from one basis to the other $\nu_{\alpha}=\sum_{i}V_{\alpha i}\nu_{i}^{\prime}$
($i=1,\text{\ensuremath{\cdots}},5$, $\alpha=e,\mu,\tau,E,F$) we
have $V_{\alpha4}=N_{\alpha}=y_{\alpha}/\sqrt{\sum_{\beta}y_{\beta}^{2}}$,
$V_{\alpha5}=N_{\alpha}^{\prime}=y_{\alpha}^{\prime}/\sqrt{\sum_{\beta}y_{\beta}^{\prime2}}$,
and $\sum_{\beta}N_{\beta}N_{\beta}^{\prime}=0$. The rest of the
elements in $V_{\alpha i}$ can be found by using the orthogonality
of $V$, which gives us $12$ equations ($9$ orthogonality and $3$
normalization conditions, because $N_{\alpha}$ and $N_{\alpha}^{\prime}$
are already normalized and orthogonal), therefore we still can choose
at will three elements of $V_{\alpha i}$; for instance we could choose
$V_{F1}=V_{F2}=V_{E1}=0$. In this case

\begin{equation}
V=\left(\begin{array}{ccccc}
V_{e1} & V_{e2} & V_{e3} & N_{e} & N_{e}^{\prime}\\
V_{\mu1} & V_{\mu2} & V_{\mu3} & N_{\mu} & N_{\mu}^{\prime}\\
V_{\tau1} & V_{\tau2} & V_{\tau3} & N_{\tau} & N_{\tau}^{\prime}\\
0 & V_{E2} & V_{E3} & N{}_{E} & N_{E}^{\prime}\\
0 & 0 & V_{F3} & N_{F} & N_{F}^{\prime}\end{array}\right)\ .\label{eq:generalrotation}\end{equation}
 Moreover, since $\sum_{\alpha}N_{\alpha}N_{\alpha}^{\prime}=0$,
the $4\times4$ mass matrix of $\nu_{4}^{\prime},\nu_{4R},\nu_{5}^{\prime}$
and $\nu_{5R}$ is block-diagonal and can be separated in two $2\times2$
matrices (for $\nu_{4}^{\prime}$ and $\nu_{4R}$ and $\nu_{5}^{\prime}$
and $\nu_{5R}$ respectively) with the same form found in the four-generation
case. Its diagonalization leads to four Majorana massive fields:\[
\nu_{a}=i\cos\theta_{a}(-\nu_{a}^{\prime}+\nu_{a}^{\prime c})+i\sin\theta_{a}(\nu_{aR}-\nu_{aR}^{c}),\]
\[
\nu_{\bar{a}}=-\sin\theta_{a}(\nu_{a}^{\prime}+\nu_{a}^{\prime c})+\cos\theta_{a}(\nu_{aR}+\nu_{aR}^{c}),\]
\begin{equation}
m_{a,\bar{a}}=\cfrac{1}{2}\left(\sqrt{m_{aR}^{2}+4m_{aD}^{2}}\mp m_{aR}\right),\label{eq:nu45masses}\end{equation}
with $a=4,5$, $\tan^{2}\theta_{a}=m_{a}/m_{\bar{a}}$, $m_{4D}=v\sqrt{\sum_{\text{\ensuremath{\alpha}}}y_{\alpha}^{2}}$
and $m_{5D}=v\sqrt{\sum_{\text{\ensuremath{\alpha}}}y_{\alpha}^{\prime2}}$
.

\subsection{Two-loop neutrino masses}

As in the case of four generations, the diagrams of figure~\ref{fig:TwoW}
(now with the four massive neutrinos running in the loop) will generate
a non-vanishing mass matrix for the three neutrinos $\nu_{\text{1}}^{\prime},\nu_{\text{2}}^{\prime},\nu_{\text{3}}^{\prime}$
given by\begin{equation}
M_{ij}=-\frac{g^{4}}{m_{W}^{4}}\sum_{a=4,5}m_{aR}m_{aD}^{2}\sum_{\text{\ensuremath{\alpha}}}V_{\alpha i}V_{\alpha a}m_{\alpha}^{2}\sum_{\beta}V_{\beta j}V_{\beta a}m_{\beta}^{2}I_{\alpha\beta}^{(a)},\label{eq:masasligeros5gen}\end{equation}
 with $I_{\alpha\beta}^{(a)}$ given by \eqref{eq:integral} with
$a$ labeling the contribution of the $4^\mathrm{th}$ and $5^\mathrm{th}$ 
generations.

To analyze this mass matrix first we will impose several phenomenological
constraints: 
\begin{description}
\item [{a)}] The model should be compatible with the observed universality
of fermion couplings and have small rates of lepton flavour violation
in the charged sector. This requires $y_{e},y_{\mu},y_{\tau},y_{e}^{\prime},y_{\mu}^{\prime},y_{\tau}^{\prime}\ll y_{E},y_{F},y_{E}^{\prime},y_{F}^{\prime}$. 
\item [{b)}] The model should fit the observed pattern of masses and mixings.
A good starting point would be to have expressions able to reproduce
the tribimaximal (TBM) mixing structure. 
\end{description}
The constraint \textbf{a) }together with the orthogonality condition
implies that $y_{E}y_{E}^{\prime}+y_{F}y_{F}^{\prime}\approx0$, which
can be satisfied, for instance, if $y_{F}=y_{E}^{\prime}=0$, that
is, $\nu_{E}$ only couples to $\nu_{4R}$ and $\nu_{F}$ only couples
to $\nu_{5R}$. Then, one can define $y_{\alpha}=y_{E}(\epsilon_{e},\epsilon_{\mu},\epsilon_{\tau},1,0)$,
$y_{\alpha}^{\prime}=y_{F}^{\prime}(\epsilon_{e}^{\prime},\epsilon_{\mu}^{\prime},\epsilon_{\tau}^{\prime},0,1)$,
where $\epsilon_{i}$ and $\epsilon_{i}^{\prime}$ are at least $\mathcal{O}(10^{-2}$)
in order to satisfy universality constraints%
\footnote{This pattern of couplings can easily be enforced by using a discrete
symmetry which is subsequently broken at order $\epsilon$. %
} (see section~\ref{sub:Universality-bounds} for more details). Thus,
to order $\epsilon$, $N_{\alpha}\approx(\epsilon_{e},\epsilon_{\mu},\epsilon_{\tau},1,0)$,
$N_{\alpha}^{\prime}\approx(\epsilon_{e}^{\prime},\epsilon_{\mu}^{\prime},\epsilon_{\tau}^{\prime},0,1)$,
and since for $i\not=4,5$ $\sum_{\alpha}V_{\alpha i}N_{\alpha}=\sum_{\alpha}V_{\alpha i}N_{\alpha}^{\prime}=0$,
all the entries $V_{\alpha i}$ with $\alpha=e,\mu,\tau$, $i=1,2,3$
can be order one. Now if we choose $V_{F1}=V_{F2}=V_{E1}=0$ one can
see that $V_{E2},V_{E3}$ are $\mathcal{O}(\text{\ensuremath{\epsilon}) while \ensuremath{V_{F3}}}$
is $\mathcal{O}(\text{\ensuremath{\epsilon}}^{\prime})$.

A further simplification occurs if we assume that $V_{E3}=0$, since
in that case $E$ only couples to $\nu_{2}^{\prime}$ and and $F$
only couples to $\nu_{3}^{\prime}$. Then, in the limit $m_{e}=m_{\mu}=m_{\tau}=0$
the neutrino mass matrix $M_{ij}$ in eq. (\ref{eq:masasligeros5gen})
is already diagonal and we can easily estimate the size of the two
larger eigenvalues by neglecting the masses of the known charged leptons
in front of $m_{E}$ and $m_{F}$. We find%
\footnote{Note that the position of the eigenvalues in $M_{ij}$ depends on
the position of the zeros in $V_{\alpha i}$. The choice we made is
very convenient to reproduce the normal hierarchy spectrum.%
}\begin{equation}
M_{22}\sim\epsilon^{2}m_{4R}\cfrac{g^{4}m_{4D}^{2}m_{E}^{2}}{m_{W}^{4}(4\pi)^{4}2}\ln\cfrac{m_{E}}{m_{\bar{4}}}\:,\quad M_{33}\sim\epsilon^{\prime2}m_{5R}\cfrac{g^{4}m_{5D}^{2}m_{F}^{2}}{m_{W}^{4}(4\pi)^{4}2}\ln\cfrac{m_{F}}{m_{\bar{5}}}\ .\label{eq:m2m3-estimate}\end{equation}

Taking $m_{F}\sim m_{5D}\sim m_{W}/g$ and $\epsilon^{\prime}\sim10^{-2}$
we find $M_{33}\sim2\times10^{-9}m_{5R}$, therefore, to obtain $M_{33}\sim0.05\,\mathrm{eV}$
we need $m_{5R}\sim20\,\mathrm{MeV}$ (or $\epsilon^{\prime}\lesssim10^{-3}$
for $m_{5R}\sim1\,\mathrm{GeV}$). Since $m_{5R}$ and $m_{4R}$ control
the splitting between the two heavy Majorana neutrinos we are naturally
in the pseudo-Dirac regime unless the $\epsilon$'s are below $10^{-4}$.
We also see that the higher $m_{4D(5D)}$ or $m_{E(F)}$, the lower
the $\epsilon$ ($\epsilon'$) that is needed for a given $m_{4R(5R)}$.
On the other hand, it is clear that the required hierarchy between
$M_{33}$ and $M_{22}$ can be easily achieved both in the normal
and the inverted hierarchy cases, while the degenerate case cannot
be fitted within this scheme since the third neutrino mass is proportional
to $m_{\tau}^{4}$. After discussing the phenomenology of the model
with more detail in section~\ref{pheno}, we present the allowed regions
of the parameter space in figure~\ref{fig:Parameter-space}.

Now let us turn to constraint \textbf{b)}, that is, the light neutrino
mixings. With our simplifying choices the diagonal entries of the
light neutrino mass matrix are proportional to $m_{\tau}^{4}$, $m_{E}^{4}$,
$m_{F}^{4}$, whereas the off-diagonal ones are proportional to $m_{\tau}^{2}m_{E}^{2}$
and $m_{\tau}^{2}m_{F}^{2}$. Therefore the neutrino states $\nu_{\text{1}}^{\prime},\nu_{\text{2}}^{\prime},\nu_{\text{3}}^{\prime}$
are very close to being the true mass eigenstates and the first $3\times3$
elements of $V$, $V_{\alpha i}$, with $\alpha=e,\mu,\tau\,,i=1,2,3$
give us directly the $\mathrm{PMNS}$ mixing matrix (up to permutations).
Then, by using the orthogonality conditions it is easy to find the
structure of Yukawas that reproduce a given pattern for the PMNS matrix.
Let us study separately the two phenomenologically viable cases, normal
hierarchy (NH) and inverted hierarchy (IH).

\subsubsection{Normal hierarchy}

In the normal hierarchy case ($m_{1}<m_{2}<m_{3}$), the experimental
data tell us that $m_{3}\approx\sqrt{\left|\bigtriangleup m_{31}^{2}\right|}\approx0.05$~eV,
$m_{2}\approx\sqrt{\bigtriangleup m_{21}^{2}}\approx0.01$~eV and
allow for $m_{1}\ll m_{2}$. The structures we have found (by choosing
$V_{F1}=V_{F2}=V_{E1}=V_{E3}=0$) automatically fall in this scheme,
since (for $m_{E,F}\gg m_{4,\bar{4},5,\bar{5}}\gg m_{W}$) we obtain
$(m_{1},m_{2},m_{3})\propto(m_{\tau}^{4}/m_{\bar{4}}^{2},m_{E}^{2},m_{F}^{2})$.
Is there any choice of the Yukawa couplings $y_{\alpha}$ and $y_{\alpha}^{\prime}$
that leads naturally to some phenomenologically successful structure,
for instance TBM? If we impose TBM in $V_{\alpha i}$ ($\alpha=e,\mu,\tau$,
$i=1,2,3$), given the structure of $N_{\alpha}$ and $N_{\alpha}^{\prime}$,
the orthogonality of $V$ (at order $\epsilon^{2}$) immediately tells
us that $\epsilon_{e}=\epsilon_{\mu}=-\epsilon_{\tau}\equiv\epsilon$,
$\epsilon_{e}^{\prime}=0,\epsilon_{\mu}^{\prime}=\epsilon_{\tau}^{\prime}\equiv\epsilon^{\prime}$,
and finally $V_{E2}=-\epsilon\sqrt{3}$, $V_{F3}=-\epsilon^{\prime}\sqrt{2}$.
Therefore, a successful choice of the Yukawas will be \begin{eqnarray}
y_{\text{\ensuremath{\alpha}}} & = & y_{E}(\epsilon,\epsilon,-\epsilon,1,0),\nonumber \\
y_{\alpha}^{\prime} & = & y_{F}^{\prime}(0,\epsilon^{\prime},\epsilon^{\prime},0,1),\label{eq:ysNH}\end{eqnarray}
 which, keeping only terms up to order $\epsilon^{2}$, leads to

\begin{eqnarray}
V & \approx & \left(\begin{array}{ccccc}
\sqrt{\cfrac{2}{3}} & \cfrac{1}{\sqrt{3}}-\cfrac{\sqrt{3}}{2}\,\epsilon^{2} & 0 & \epsilon & 0\\
-\cfrac{1}{\sqrt{6}} & \cfrac{1}{\sqrt{3}}-\cfrac{\sqrt{3}}{2}\,\epsilon^{2} & \cfrac{1}{\sqrt{2}}-\cfrac{1}{\sqrt{2}}\,\epsilon^{\prime2} & \epsilon & \epsilon^{\prime}\\
\cfrac{1}{\sqrt{6}} & -\cfrac{1}{\sqrt{3}}+\cfrac{\sqrt{3}}{2}\,\epsilon^{2} & \cfrac{1}{\sqrt{2}}-\cfrac{1}{\sqrt{2}}\,\epsilon^{\prime2} & -\epsilon & \epsilon^{\text{\ensuremath{\prime}}}\\
0 & -\epsilon\,\sqrt{3} & 0 & 1-\cfrac{3}{2}\,\epsilon^{2} & 0\\
0 & 0 & -\epsilon^{\prime}\,\sqrt{2} & 0 & 1-\epsilon^{\prime2}\end{array}\right)+\mathrm{\mathcal{O}(\epsilon^{3})}.\label{eq:rotationNH}\end{eqnarray}
 Assuming that $m_{E,F}\gg m_{4,\bar{4},5,\bar{5}}\gg m_{W}$, we
find: \begin{equation}
m_{2}=-\cfrac{3g^{4}}{m_{W}^{4}}\epsilon^{2}m_{4D}^{2}m_{4R}m_{E}^{4}I_{EE}\approx\cfrac{3g^{4}}{2(4\pi)^{4}m_{W}^{4}}\epsilon^{2}m_{4D}^{2}m_{4R}m_{E}^{2}\ln\cfrac{m_{E}}{m_{\bar{4}}},\label{eq:m2-NH}\end{equation}
 \begin{equation}
m_{3}=-\cfrac{2g^{4}}{m_{W}^{4}}\epsilon^{\prime2}m_{5D}^{2}m_{5R}m_{F}^{4}I_{FF}\approx\cfrac{g^{4}}{(4\pi)^{4}m_{W}^{4}}\epsilon^{\prime2}m_{5D}^{2}m_{5R}m_{F}^{2}\ln\cfrac{m_{F}}{m_{\bar{5}}}\ ,\label{eq:m3-NH}\end{equation}
and the required ratio $m_{3}/m_{2}\approx5$ can be easily accommodated,
for instance if the fifth generation is heavier than the fourth one
or $\epsilon^{\prime}>\epsilon$.

\subsubsection{Inverted hierarchy}

In the inverted hierarchy case ($m_{3}<m_{1}\lesssim m_{2}$), we
have $m_{2}\approx m_{1}\approx\sqrt{\left|\bigtriangleup m_{31}^{2}\right|}\approx0.05$~eV
and $m_{3}\ll m_{1}$ is allowed. Therefore now we need $(m_{1},m_{2},m_{3})\propto(m_{E}^{2},m_{F}^{2},m_{\tau}^{4}/m_{\bar{4}}^{2})$,
which is just a cyclic permutation of the three eigenvalues. This
ordering cannot be obtained directly with our previous choice for
the zeroes in $V_{\alpha i}$, so now it is convenient to choose $V_{F1}=V_{F3}=V_{E3}=V_{E2}=0$
instead. Following the same procedure as above, we find that

\begin{eqnarray}
y_{\alpha} & = & y_{E}(-2\epsilon,\epsilon,-\epsilon,1,0),\nonumber \\
y_{\alpha}^{\prime} & = & y_{F}^{\prime}(\epsilon^{\prime},\epsilon^{\prime},-\epsilon^{\prime},0,1)\label{eq:ysIH}\end{eqnarray}
 will reproduce the desired TBM pattern, leading to

\begin{equation}
V\approx\left(\begin{array}{ccccc}
\sqrt{\cfrac{2}{3}}-\sqrt{6}\epsilon^{2} & \cfrac{1}{\sqrt{3}}-\cfrac{\sqrt{3}}{2}\,\epsilon^{\prime2} & 0 & -2\epsilon & \epsilon^{\prime}\\
-\cfrac{1}{\sqrt{6}}+\sqrt{\frac{3}{2}}\,\epsilon^{2} & \cfrac{1}{\sqrt{3}}-\cfrac{\sqrt{3}}{2}\,\epsilon^{\prime2} & \cfrac{1}{\sqrt{2}} & \epsilon & \epsilon^{\prime}\\
\cfrac{1}{\sqrt{6}}-\sqrt{\frac{3}{2}}\,\epsilon^{2} & -\cfrac{1}{\sqrt{3}}+\cfrac{\sqrt{3}}{2}\,\epsilon^{\prime2} & \cfrac{1}{\sqrt{2}} & -\epsilon & -\epsilon^{\text{\ensuremath{\prime}}}\\
\epsilon\,\sqrt{6} & 0 & 0 & 1-3\,\epsilon^{2} & 0\\
0 & -\epsilon^{\prime}\,\sqrt{3} & 0 & 0 & 1-\cfrac{3}{2}\,\epsilon^{\prime2}\end{array}\right)+\mathrm{\mathcal{O}(\epsilon^{3})}\ .\label{eq:rotationIH}\end{equation}
 Assuming that $m_{E,F}\gg m_{4,\bar{4},5,\bar{5}}\gg m_{W}$, we
get:\begin{equation}
m_{1}\approx\cfrac{3g^{4}}{(4\pi)^{4}m_{W}^{4}}\,\epsilon^{2}m_{4D}^{2}m_{4R}m_{E}^{2}\ln\cfrac{m_{E}}{m_{\bar{4}}},\label{eq:m1-IH}\end{equation}
 \begin{equation}
m_{2}\approx\cfrac{3g^{4}}{2(4\pi)^{4}m_{W}^{4}}\epsilon{}^{\prime2}m_{5D}^{2}m_{5R}m_{F}^{2}\ln\cfrac{m_{F}}{m_{\bar{5}}}\ ,\label{eq:m2-IH}\end{equation}
while the ratio of masses between the heaviest neutrinos, $m_{1}/m_{2}\approx1$
can be obtained by choosing the different parameters in their natural
range.

Thus, the model accommodates the light neutrino masses and mixings.
In the next section we will analyse current phenomenological bounds
on the mixings between the new generations and the first three, $\epsilon,\epsilon'$.

\subsection{The parameters of the model\label{pheno}}

We have seen above that neutrino masses are proportional to $\epsilon^{2}m_{4R}$
(or $\epsilon^{\prime2}m_{5R}$) and a product of masses, $m_{4D}^{2}m_{E}^{2}$
(or $m_{5D}^{2}m_{F}^{2})$, which come from the Higgs mechanism and
are proportional to Yukawa couplings. As discussed in section~\ref{sec:frame}
the values of these masses cannot vary too much; perturbative unitarity
requires they are smaller than about $1\,\mathrm{TeV}$~\cite{Chanowitz:1978mv}
while lower limits for charged leptons masses from colliders are about
$100\,\mathrm{GeV}$. Lower limits for neutral fermions are a bit
less uncertain. In the case of unstable pure Dirac neutrinos ($m_{aR}=0$,
$a=4,5$) the neutrino masses are basically $m_{aD}$ and the lower
limits are about $90\,\mathrm{GeV}$, therefore, in that case, $m_{aD}\gtrsim90\,\mathrm{GeV}$.
If neutrinos have both Dirac and Majorana mass terms ($m_{aR}\not=0$)
the masses are given by eq.~(\ref{eq:nu45masses}) and the lower
limits are%
\footnote{Notice that in our scenario there is a lower bound on the mixing $\epsilon$,
in order to obtain the correct scale of light neutrino masses, which
implies that the heavy neutrinos would have decayed inside the detector
at LEP.%
} $m_{\bar{a}}\ge m_{a}>63\,\mathrm{GeV}$, then the upper limits on
$m_{aD}<1\,\mathrm{TeV}$ automatically imply $m_{aR}\lesssim16\,\mathrm{TeV}$
and therefore $m_{\bar{a}}\lesssim16\,\mathrm{TeV}$. More generally
in figure~\ref{fig:mu-vs-md} we present the allowed regions in the
plane $m_{aR}$~vs~$m_{aD}$ given the lower bound on $m_{a}>63\,\mathrm{GeV}$
and the upper limit on $m_{aD}<1000\,\mathrm{GeV}.$ We also plot
the lines corresponding to $m_{a}=200$, $400$, $600$ and $800\,\mathrm{GeV}$.
\FIGURE{
\begin{centering}
\includegraphics[scale=0.5]{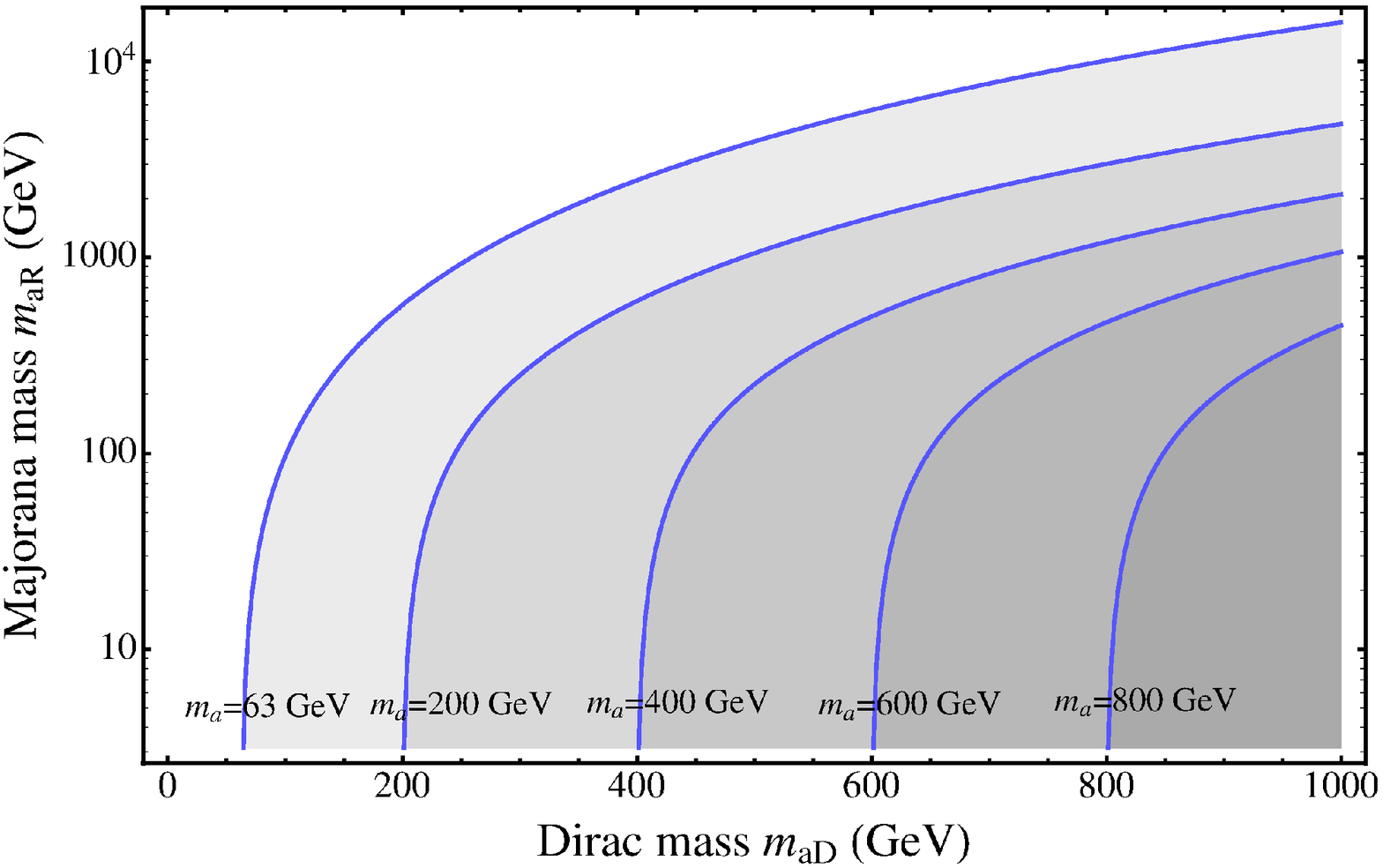}
\par\end{centering}

\caption{Allowed region in $m_{aR}$--$m_{aD}$ given the present lower limit
($m_{a}>63\,\mathrm{GeV}$) on the mass of an extra generation neutrino
($a=4,5$ refers the 4$^{\mathrm{th}}$ or 5$^{\mathrm{th}}$ generation).\label{fig:mu-vs-md}}
}

To be definite we will take\[
100\,\mathrm{GeV}<m_{4D},m_{5D},m_{E},m_{F}<1000\,\mathrm{GeV}\,,\]
\[
63\,\mathrm{GeV\lesssim}m_{a}\leq m_{aD}\leq m_{\bar{a}}\lesssim16\,\mathrm{TeV\,,}\]
with $m_{a}m_{\bar{a}}=m_{aD}^{2}$ and $m_{\bar{a}}-m_{a}=m_{aR}$.
In addition there are strong constrains from the electroweak oblique
parameters which in the pure Dirac case require some degeneracy of
masses, $m_{4D}\simeq m_{E}$ ($m_{5D}\simeq m_{F}$). However, these
constraints depend on the complete spectrum of the theory (masses
of quarks and leptons from new generations and the Higgs boson mass)
and are less certain. In fact, contributions from the splitting of
masses in the quark sector can be compensated in part by lepton contributions
with large $m_{aR}$~\cite{Bertolini:1990ek,Kniehl:1992ez}, which,
if we do take into account the constraints set by LEP II
can vary from essentially zero (Dirac case) to $16\,\mathrm{TeV}$. 

The other parameters that enter neutrino masses are the $\epsilon$'s,
which characterize the mixing of light neutrinos with heavy neutrinos,
and the $m_{R}$'s, which characterize the amount of total lepton
number breaking. The $\epsilon$ parameters will produce violations
of universality and flavour lepton number conservation in low energy
processes. The combination of data from these processes will allow
us to constrain both $\epsilon$ and $\epsilon^{\prime}$. On the
other hand, to obtain information on the $m_{R}$'s we will use the
light neutrino masses, which in the model are Majorana particles.
We will also study the contributions of the heavy neutrinos to neutrinoless
double beta decay.

\subsubsection{Lepton flavour violation processes ($\mu\rightarrow e\gamma$ and
$\mu$--$e$ conversion)}

The general expression for the branching ratio of $\mu\rightarrow e\gamma$
produced through a virtual pair $W$-neutrino is: \begin{equation}
B(\mu\rightarrow e\gamma)=\frac{3\alpha}{2\pi}\,\left|\delta_{\nu}\right|^{2},\label{muegamma-general}\end{equation}
 where \begin{equation}
\delta_{\nu}=\sum_{i}\, U_{ei}U_{\mu i}^{*}\, H\left(\nicefrac{m_{\chi_{i}}^{2}}{m_{W}^{2}}\right)\label{eq:delta-nu}\end{equation}
and $H$ is the loop function for this process~\cite{Buras:2010cp}
\[
H(x)=\frac{x\left(2x^{2}+5x-1\right)}{4(x-1)^{3}}-\frac{3x^{3}\log(x)}{2(x-1)^{4}},\]
with $m_{\chi_{i}}$ the masses of all heavy neutrinos running in
the loop and $U_{ei}$ and $U_{\mu i}$ their couplings to the electron
and the muon respectively. In \eqref{muegamma-general} we have used
the unitarity of the mixing matrix and neglected the light neutrino
masses to rewrite the final result only in terms the heavy neutrino
contributions. Then, as the mixings of the heavy neutrinos with the
light leptons are different in normal and inverted hierarchy, so are
the $\mu\rightarrow e\gamma$ amplitudes generated; one can see just
by inspection of the mixing matrices, eqs.~(\ref{eq:rotationNH}) and (\ref{eq:rotationIH})
that in NH only the pair $\nu_{4},\nu_{\bar{4}}$ couples to both
the electron and the muon, whereas in IH the four heavy neutrinos
contribute to the process. The predicted branching ratios are \begin{align}
\text{NH:}\qquad & B(\mu\rightarrow e\gamma)=\frac{3\alpha}{2\pi}\,\bar{H}_{4}^{2}\,\epsilon^{4},\label{eq:BRmu2egamma-NH}\\
\text{IH:}\qquad & B(\mu\rightarrow e\gamma)=\frac{3\alpha}{2\pi}\,\left[\bar{H}_{5}\,\epsilon^{\prime2}-2\,\bar{H}_{4}\,\epsilon^{2}\right]^{2}\,,\label{eq:BRmu2egamma-IH}\end{align}
where \[
\bar{H}_{a}\equiv\cos^{2}\theta_{a}H(m_{a}^{2}/m_{W}^{2})+\sin^{2}\theta_{a}H(m_{\bar{a}}^{2}/m_{W}^{2})\,.\]
Now since $H(x)$ is a monotonically increasing function and $m_{\bar{a}}\geq m_{a}>63\,\mathrm{GeV}$
we have $\bar{H}_{a}\ge H(m_{a}^{2}/m_{W}^{2})>0.09$ which gives
the less stringent constraint on $\epsilon$ and $\epsilon^{\prime}$.
The experimental bound reads $B(\mu\rightarrow e\gamma)<1.2\times10^{-11}$,
and it is translated into \begin{alignat}{2}
\mathrm{NH:}\qquad & \epsilon & \,<\, & 0.03,\label{eq:eps-limits-nu2eg-NH}\\
\text{IH:}\qquad & |\epsilon^{\prime2}-2\,\epsilon^{2}| & \,<\, & 7\times10^{-4}.\label{eq:eps-limits-mu2eg-IH}\end{alignat}

\FIGURE{
\begin{centering}
\includegraphics[scale=0.6]{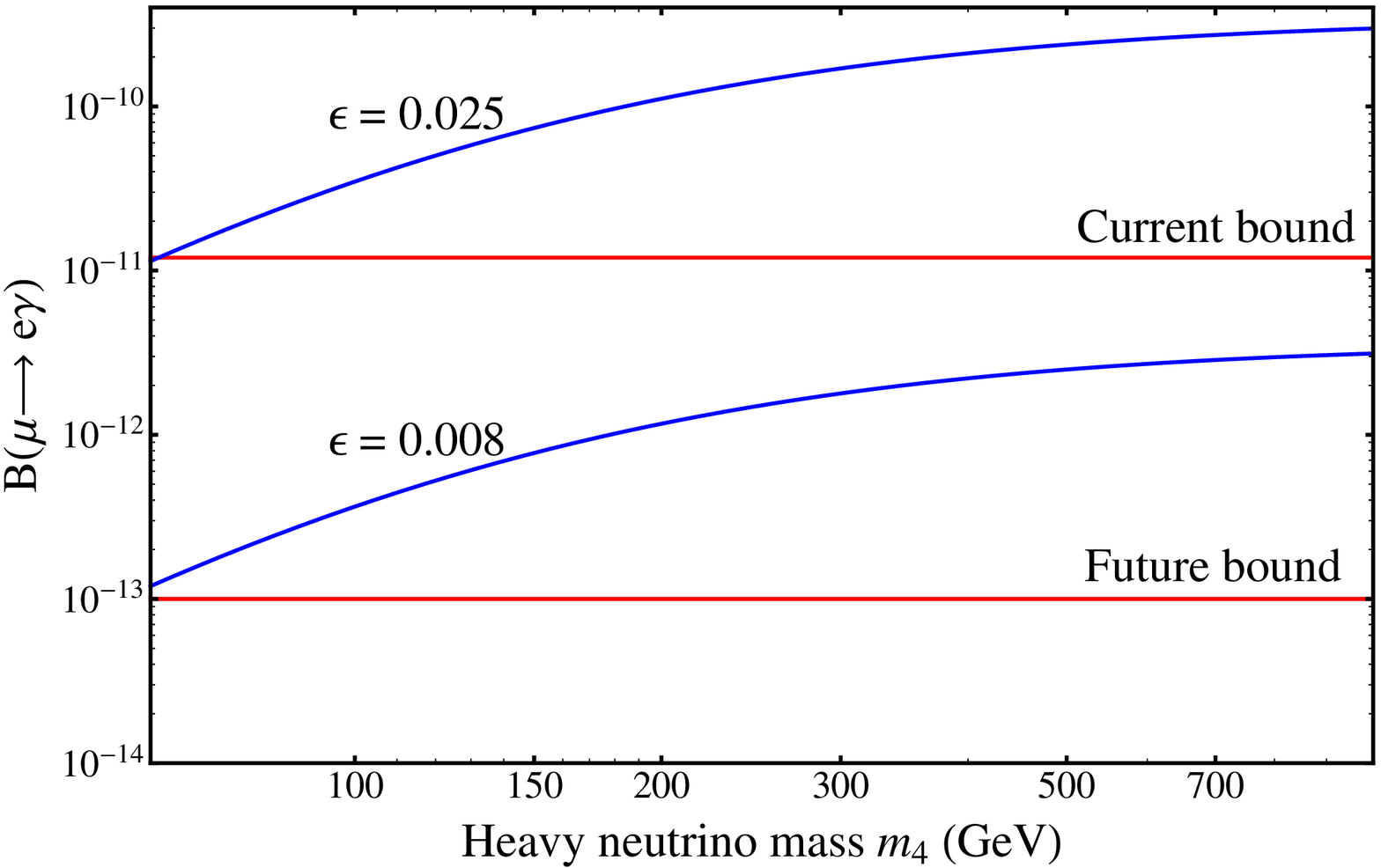} 
\par\end{centering}

\centering{}\caption{$B(\mu\rightarrow e\gamma)$ against $m_{4}$ 
for different values
of $\epsilon$ in the NH case. We also display present and future
limits on $B(\mu\rightarrow e\gamma)$ as horizontal lines.
\label{fig:mu2egamma}}
}

To see how these bounds depend on the masses of the heavy neutrinos
we display in figure~\ref{fig:mu2egamma} $B(\mu\rightarrow e\gamma)$
against the mass of the heavy neutrino $m_{4}$ in the NH case. For
IH we expect similar results unless there are strong cancellations.
We also display as horizontal lines present limits \cite{Nakamura:2010zzi}
$B(\mu\rightarrow e\gamma)<1.2\times10^{-11}$ and near future limits
\cite{Cei:2010zz}. From the figure we can extract a conservative
bound of the order of the one quoted above, $\epsilon<0.03$. 

Some extra information could be extracted from $\tau\rightarrow e\gamma$
and $\tau\rightarrow\mu\gamma$. Thus, from $B(\tau\rightarrow e\gamma)<3.3\times10^{-8}$
we obtain $\epsilon<0.3$ in the case of NH and $|\epsilon^{\prime2}-2\,\epsilon^{2}|\,<\,0.08,$
limits that show exactly the same dependence on $\epsilon$ and $\epsilon^{\prime}$
as the one obtained in $\mu\rightarrow e\gamma$, but which are roughly
one order of magnitude worse. From $B(\tau\rightarrow\mu\gamma)<4.4\times10^{-8}$,
although the bounds are of the same order of magnitude as for $B(\tau\rightarrow e\gamma)$,
we obtain different combinations of $\epsilon$'s, $|\epsilon^{\prime2}-\epsilon^{2}|<0.09$
for NH and $\epsilon^{\prime2}+\epsilon^{2}<0.09$ for IH.

Another very interesting process which gives information on $\epsilon$
is $\mu$--$e$ conversion in nuclei. From present data \cite{Nakamura:2010zzi}
one obtains bounds similar to the limit obtained from $\mu\rightarrow e\gamma$.
However, there are plans to improve the sensitivity in $\mu$--$e$
conversion in $4$ and even $6$ orders of magnitude \cite{Hungerford:2009zz},
therefore we expect much stronger bounds in the future coming from
$\mu$--$e$ conversion. Strong correlations between both processes
exist, as can be seen in \cite{Buras:2010cp}.

\subsubsection{Universality bounds\label{sub:Universality-bounds}}

New heavy generations that couple to the observed fermions can potentially
lead to violations of universality in charged currents because of
the ``effective'' lack of unitarity in the mixings when the heavy
generations cannot be produced. Data from neutrino oscillation experiments
can also be used to constrain deviations from unitarity of some of
the elements of the leptonic mixing matrix \cite{Antusch:2006vwa}, however, 
in our scenario they lead to weaker bounds than the ones obtained here.

There are different types of universality bounds which constrain the
mixings of light fermions with new generations: 
\begin{itemize}
\item Lepton-hadron universality. One compares weak couplings of quarks
and leptons using muon decay and nuclear $\beta$ decay, which are
very well tested. In our case this involves mixings both in the quark
and lepton sectors and they are not useful to test individually the
lepton mixings we are interested in. 
\item Relations between muon decay, $m_{Z}$, $m_{W}$ and the weak mixing
angle $\sin^{2}\theta_{W}$. These are very well-determined relations
in the SM, and in our case they are modified because the heavy neutrinos
cannot be produced in ordinary muon decay. Unfortunately, these relations
depend strongly on the $\rho$ parameter, which receives contributions
from the Higgs and very large contributions from the heavy fermions
of the new generations. Therefore, although these type of relations
could be used to set bounds on the $\epsilon$'s, they would depend
on other unknown parameters. 
\item Ratios of decay widths of similar processes. The bounds obtained from
this type of processes are very robust because most of the uncertainties
cancel in the ratios. We will only consider the most precise among
these ratios, which are well measured and can be computed accurately
\cite{Pich:2008ni,Cirigliano:2007ga,Marciano:1988vm}: \begin{equation}
R_{\pi\rightarrow e/\pi\rightarrow\mu}\equiv\frac{\Gamma(\pi\rightarrow e\bar{\nu})}{\Gamma(\pi\rightarrow\mu\bar{\nu})},\label{eq:Rpiemu}\end{equation}
 \begin{equation}
R_{\tau\rightarrow e/\tau\rightarrow\mu}\equiv\frac{\Gamma(\tau\rightarrow e\bar{\nu}\nu)}{\Gamma(\tau\rightarrow\mu\bar{\nu}\nu)},\label{eq:Rtauemu}\end{equation}
 \begin{equation}
R_{\tau\rightarrow e/\mu\rightarrow e}\equiv\frac{\Gamma(\tau\rightarrow e\bar{\nu}\nu)}{\Gamma(\mu\rightarrow e\bar{\nu}\nu)}=B_{\tau\rightarrow e}\frac{\tau_{\mu}}{\tau_{\tau}},\label{eq:Rtaumu}\end{equation}
 \begin{equation}
R_{\tau\rightarrow\mu/\mu\rightarrow e}\equiv\frac{\Gamma(\tau\rightarrow\mu\bar{\nu}\nu)}{\Gamma(\mu\rightarrow e\bar{\nu}\nu)}=B_{\tau\rightarrow\mu}\frac{\tau_{\mu}}{\tau_{\tau}},\label{eq:Rtaue}\end{equation}

\end{itemize}
where $B_{\tau\rightarrow f}=\Gamma(\tau\rightarrow f\bar{\nu}\nu)/\Gamma(\tau\rightarrow\mathrm{all})$
is the branching ratio of the tau decay to the fermion $f$, and $\tau_{f}=1/\Gamma(f\rightarrow\mathrm{all})$
its lifetime. In our model there are corrections to these ratios because
$\nu_{e}$, $\nu_{\mu}$ and $\nu_{\tau}$ have a small part of $\nu_{4,\bar{4}}$
and $\nu_{5,\bar{5}}$, which are heavy and cannot be produced. This
leads to an additional violation of universality which depends on
the mixings of $\nu_{e}$, $\nu_{\mu}$ and $\nu_{\tau}$ with $\nu_{4,\bar{4}}$
and $\nu_{5,\bar{5}}$. For $R_{\pi\rightarrow e/\pi\rightarrow\mu}$,
and using the $V_{\alpha i}$ in \eqref{eq:rotationNH} and \eqref{eq:rotationIH},
we find that \begin{equation}
\dfrac{R_{\pi\rightarrow e/\pi\rightarrow\mu}}{R_{\pi\rightarrow e/\pi\rightarrow\mu}^{\mathrm{SM}}}=\frac{\left|V_{e1}\right|^{2}+\left|V_{e2}\right|^{2}+\left|V_{e3}\right|^{2}}{\left|V_{\mu1}\right|^{2}+\left|V_{\mu2}\right|^{2}+\left|V_{\mu3}\right|^{2}}=\left\{ \begin{array}{lcl}
1+\epsilon^{\prime2} &  & \mathrm{NH}\\
1-3\epsilon^{2} &  & \mathrm{IH}\end{array}\right.\,.\label{eq:limitRemu}\end{equation}

$R_{\tau\rightarrow e/\tau\rightarrow\mu}/R_{\tau\rightarrow e/\tau\rightarrow\mu}^{\mathrm{SM}}$
tests exactly the same couplings, therefore the result is the same
as in \eqref{eq:limitRemu}. 

$R_{\tau\rightarrow e/\mu\rightarrow e}$ gives a different information
because it tests $\tau/\mu$ universality; however, we find that for
both NH and IH $R_{\tau\rightarrow e/\mu\rightarrow e}/R_{\tau\rightarrow e/\mu\rightarrow e}^{\mathrm{SM}}=1$,
and since it is independent of the $\epsilon$'s, this process does
not give any further information. This is a consequence of our choice
for the Yukawa couplings%
\footnote{Which, in turn, is a consequence of the TBM structure we wanted to
reproduce.%
}, which, up to signs, are equal for the $\tau$ and $\mu$ neutrinos.

Finally for $R_{\tau\rightarrow\mu/\mu\rightarrow e}$, using our
mixing matrices, we find \begin{equation}
\dfrac{R_{\tau\rightarrow\mu/\mu\rightarrow e}}{R_{\tau\rightarrow\mu/\mu\rightarrow e}^{\mathrm{SM}}}=\frac{\left|V_{\text{\ensuremath{\tau}}1}\right|^{2}+\left|V_{\tau2}\right|^{2}+\left|V_{\tau3}\right|^{2}}{\left|V_{e1}\right|^{2}+\left|V_{e2}\right|^{2}+\left|V_{e3}\right|^{2}}=\left\{ \begin{array}{lcl}
1-\epsilon^{\prime2} &  & \mathrm{NH}\\
1+3\epsilon^{2} &  & \mathrm{IH}\end{array}\right.\,,
\label{eq:limitRtaue}\end{equation}
which gives exactly the inverse combinations of those obtained 
from $e/\mu$ universality tests.

\begin{center}
\TABLE{
\begin{centering}
\begin{tabular}{|c|c|c|c|}
\hline 
Observable & $R$ \cite{Nakamura:2010zzi} & $R^{\mathrm{SM}}$ & $R/R^{\mathrm{SM}}$\tabularnewline
\hline
\hline 
$R_{\pi\rightarrow e/\pi\rightarrow\mu}$ & $\left(1.230\pm0.004\right)\times10^{-4}$  & $\left(1.2352\pm0.0001\right)\times10^{-4}$  & $0.996\pm0.003$.\tabularnewline
\hline 
$R_{\tau\rightarrow e/\tau\rightarrow\mu}$  & $1.028\pm0.004$  & $1.02821\pm0.00001$  & $1.000\pm0.004$\tabularnewline
\hline 
$R_{\tau\rightarrow\mu/\mu\rightarrow e}$ & $\left(1.31\pm0.06\right)\times10^{6}$ & $\left(1.3086\pm0.0006\right)\times10^{6}$  & $1.001\pm0.004$\tabularnewline
\hline
\end{tabular}
\par\end{centering}

\caption{Relevant universality tests \label{tab:universality-tests}}
}

\par\end{center}

Therefore, if we combine the three results, $R_{\pi\rightarrow e/\pi\rightarrow\mu}/R_{\pi\rightarrow e/\pi\rightarrow\mu}^{\mathrm{SM}}$,
$R_{\tau\rightarrow e/\tau\rightarrow\mu}/R_{\tau\rightarrow e/\tau\rightarrow\mu}^{\mathrm{SM}}$
and $(R_{\tau\rightarrow\mu/\mu\rightarrow e}/
R_{\tau\rightarrow\mu/\mu\rightarrow e}^{\mathrm{SM}})^{-1}$
and use the data collected in table~\eqref{tab:universality-tests}, we obtain
\begin{equation}
0.998\pm0.002=\left\{ \begin{array}{ll}
1+\epsilon^{\prime2} & \mathrm{NH}\\
1-3\epsilon^{2} & \mathrm{IH}\end{array}\right.\,,\label{eq:universality-epsilons1}\end{equation}
 which translates into the following upper $90\%$~C.L. limits on
$\epsilon^{\prime}$ and $\epsilon$\\
 \begin{alignat}{2}
\mathrm{NH:\qquad} & \epsilon^{\prime} & \,<\,0.04,\label{eq:limit-univ-NH}\\
\mathrm{IH:\qquad} & \epsilon & \,<\,0.04.\label{eq:limit-univ-IH}\end{alignat}
Notice that although in the IH case we have more sensitivity than
in the NH case because of the factor of $3$ we finally obtain similar
limits in the two cases. This is because in the NH case the deviation
from $1$ obtained in the model is always positive while the present
measured value is slightly smaller than $1$ (in both cases we used
the Feldman~\&~Cousins prescription \cite{Feldman:1997qc} to set 90\%
C.L. limits).

Now we can use all data from LFV and universality and conclude that
in the NH case we have $\epsilon<0.03$, basically from $\mu\rightarrow e\gamma$,
and $\epsilon^{\prime}<0.04$, basically from universality tests.
In the IH case we obtain that, except in a narrow band around $\epsilon^{\prime2}\simeq2\epsilon^{2}$,
$\epsilon\lesssim0.02$ and $\epsilon^{\prime}\lesssim0.03$ basically
from $\mu\rightarrow e\gamma$; if $\epsilon^{\prime2}\simeq2\epsilon^{2}$
there is a cancellation in $\mu\rightarrow e\gamma$ but still one
can combine these data with the universality limits to obtain $\epsilon\lesssim0.04$
and $\epsilon^{\prime}\lesssim0.06$.

\subsubsection{Neutrinoless double beta decay ($0\nu2\beta$) }

As commented above $m_{4R},m_{5R}$ are the relevant parameters which
encapsulate the non-conservation of total lepton number and they control,
together with the $\epsilon$'s, the neutrino masses. Therefore, it
would be useful to have additional independent information on these
parameters. The most promising experiments to test the non-conservation
of total lepton number are neutrinoless double beta decay experiments.
The standard contribution, produced by light neutrinos, to $0\nu2\beta$
has largely been studied (for a recent review see for instance \cite{Avignone:2007fu})
and, given the expected future sensitivity, $m_{ee}=0.01\,\mathrm{eV}$,
it will be very difficult to see it unless the neutrino spectrum is
inverted or degenerate. However, if heavy neutrinos from new families
are Majorana particles, they lead to tree-level effects in neutrinoless
double beta decay \cite{Lenz:2011gd}, while, in our scenario, light neutrino 
masses are  
generated at two loops; thus, in principle, it is possible that these 
new contributions dominate over the standard ones.

The contribution of new generation heavy neutrinos (with mass larger
than about the proton mass, $m_{p}$) to the rate of neutrinoless
double beta decay can be written in terms of an effective mass, \begin{eqnarray}
\left\langle M_{N}\right\rangle ^{-1} & = & \sum_{a}U_{ea}^{2}M_{a}^{-1},\label{eq:MNteo-1}\end{eqnarray}
 where $U_{ea}$ is the coupling of the electron to the left-handed
component of the heavy neutrino $a$. The non-observation of neutrinoless
double beta decay implies that \cite{Pas:2000vn} \begin{equation}
\left\langle M_{N}\right\rangle >10^{8}\,\text{GeV}\ .\label{eq:MNbound}\end{equation}

In the case of NH the electron only couples to $\nu_{4}$ and $\nu_{\bar{4}}$
(see eqs. \eqref{eq:rotationNH} and \eqref{eq:nu45masses}), thus
\begin{equation}
\left\langle M_{N}\right\rangle ^{-1}=2\epsilon{}^{2}(\cfrac{\cos^{2}\theta}{m_{4}}-\cfrac{\sin^{2}\theta}{m_{\bar{4}}})=\epsilon^{2}\cfrac{m_{4R}}{m_{4D}^{2}}\ ,\label{eq:PDlimitonubb}\end{equation}
 and using \eqref{eq:MNbound}, we get: \begin{eqnarray}
m_{4R}\epsilon^{2}/m_{4D}^{2} & < & 10^{-8}\,\mathrm{GeV^{-1}}.\label{eq:m1epsonubb}\end{eqnarray}
This is the same combination of the relevant parameters ($m_{4R}\epsilon^{2}$)
that appears in \eqref{eq:m2-NH} for $m_{2}$, therefore we can use
neutrino data to set bounds on the heavy neutrino contribution to
neutrinoless double beta decay written in terms of the effective mass.
We obtain \begin{equation}
\left\langle M_{N}\right\rangle =\frac{3g^{4}m_{4D}^{4}m_{E}^{2}\ln\cfrac{m_{E}}{m_{\bar{4}}}}{2m_{2}(4\pi)^{4}m_{W}^{4}}\gtrsim2\times10^{11}\,\mathrm{GeV}\ ,\end{equation}
where we have used $m_{2}\sim0.01\,\mathrm{eV}$, typical values for
$m_{4D}\sim m_{E}\sim100\,\mathrm{GeV}$ and $\ln (m_{E}/m_{\bar{4}})\sim1$.
This is far from present, eq.~\ref{eq:MNbound}, and future sensitivities.

In the case of IH, the effective mass is given by \begin{equation}
\left\langle M_{N}\right\rangle ^{-1}=4\epsilon^{2}m_{4R}/m_{4D}^{2}+\epsilon^{\prime2}m_{5R}/m_{5D}^{2}\ ,\end{equation}
 leading also to unobservable effects in $0\nu2\beta$ decay.

To summarize all phenomenological constraints on the model, 
in figure~\ref{fig:Parameter-space} we show in blue 
the allowed region in the $\epsilon-m_{4R}$ plane,
which leads to $M_{33}\sim0.05\,\mathrm{eV}$ varying the charged
lepton masses $m_{E}\,(m_{F})$ and the Dirac neutrino masses $m_{4D}\,(m_{5D})$
between 100~GeV--1~TeV, and imposing the LEP bound on the physical
neutrino mass, $m_{4}>63\,\mathrm{GeV}$. We also plot the present
bounds on the mixings $\epsilon\,(\epsilon^{\prime})$ from 
$\mu\rightarrow e\gamma$
and future limits from $\mu$--$e$ conversion if expectations are
attained. 

\FIGURE{
\begin{centering}
\includegraphics[scale=0.5]{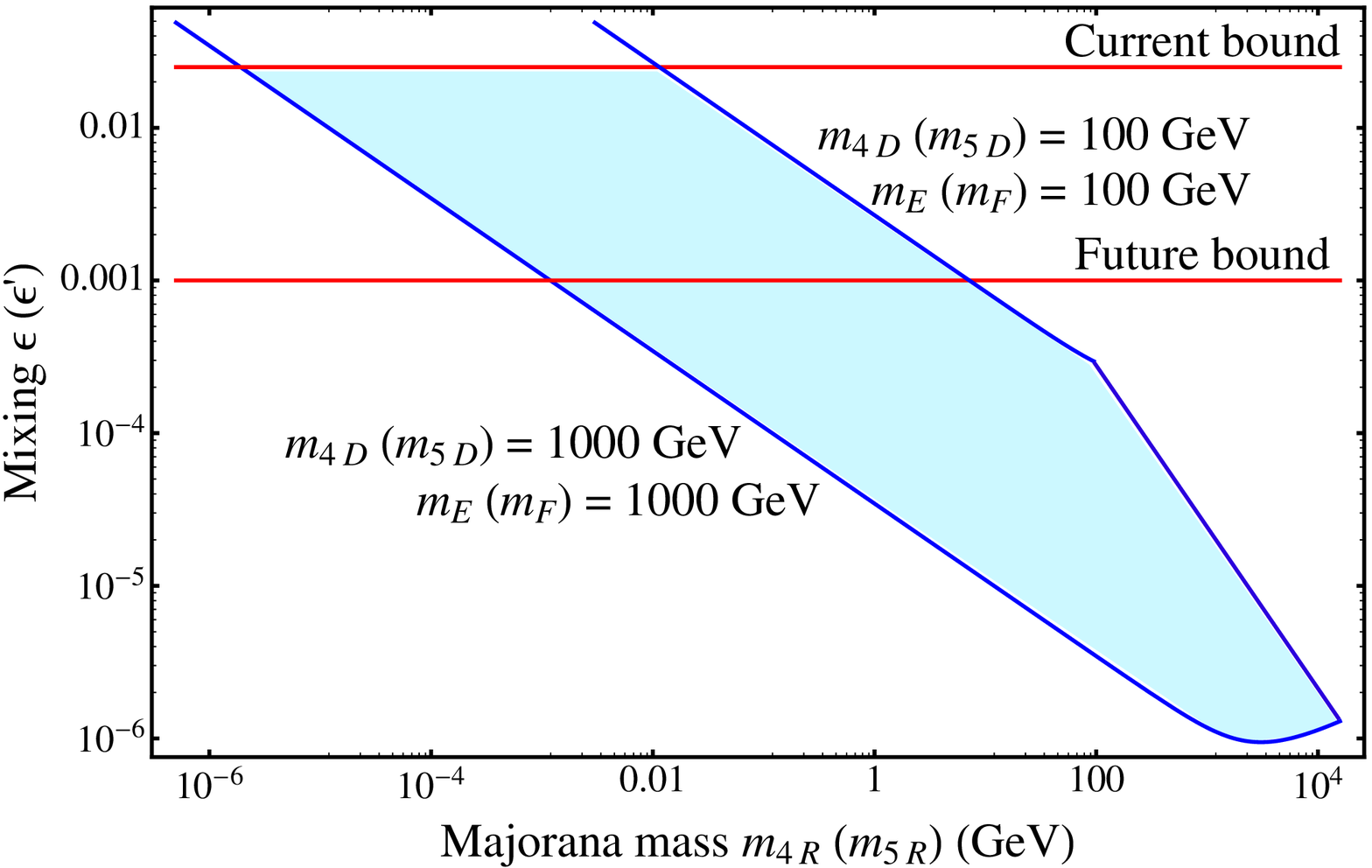}
\par\end{centering}

\caption{Parameter space that predicts the right scale for heavy and light
neutrinos (blue region between the curves). 
As a comparison we also present the current bound from
$\mu\rightarrow e\gamma$ and future limits from $\mu$--$e$ conversion
experiments. \label{fig:Parameter-space}}
}

\section{Collider signatures\label{colliders}}

As we mentioned before, the LHC offers a unique opportunity to discover
(or exclude) new sequential generations of quarks and leptons. For
instance, with 1~${\rm fb}^{-1}$ at 7~TeV, the exclusion bound on
$b'$ would reach 500 GeV via $b'\rightarrow Wt$ decay channel, close
to the partial wave unitarity bound. Even if the $t'$ and $b'$ are
too heavy to be seen directly, their effects may be manifest at LHC,
since they induce a large $gg\rightarrow ZZ$ signal \cite{Chanowitz:1995rv}.
See also \cite{Murayama:2010xb} for prospects of detecting very long-lived
fourth generation quarks, i.e., in the case of extremely small mixings
with the lighter three generations.

Regarding the lepton sector, the standard searches for a fourth generation
have to be restricted to the parameter space which leads to the correct
light neutrino mass scale, depicted in figure~\ref{fig:Parameter-space}.
The expected signatures depend on the nature (Dirac or Majorana) of
the neutrinos, which are generally assumed to be the lightest states.

Most theoretical analysis of fourth generation Majorana neutrino at
hadron colliders have focused on the process $q\bar{q}'\rightarrow W^{\pm}\rightarrow\nu_{4}\ell^{\pm}$,
where the fourth generation neutrino is produced in association with
a light charged lepton \cite{Han:2006ip,Atre:2009rg,del Aguila:2007em,delAguila:2008ir}.
Subsequently, if neutrinos are Majorana, they will decay through $\nu_{4}\rightarrow W^{\mp}\ell^{\pm}$,
leading to the low-background like-sign di-lepton signature in half
the events. However in our model the cross section for this process
is suppressed both by the mixing of the extra generations with the
first three and by the small Majorana masses $m_{4R,5R}$, much as
in the neutrinoless double beta decay discussed above, so it will
not be observable at LHC for the parameter range that reproduce the
correct scale of light neutrino masses. 

Alternatively, the lighter neutrinos can be pair-produced via an $s$-channel
$Z$ boson, $q\bar{q}\rightarrow Z\rightarrow\nu_{I}\nu_{J}$ ($I,J=4,\bar{4}$)
\cite{Rajaraman:2010ua}. Although the $W$ production has a higher
cross section than the $Z$ at hadron colliders, and the mass reach
is enhanced when only one heavy particle is produced, if the mixing
angle between the extra generations and the light ones is less than
about $10^{-6}$ the neutrino production rates in the $W$ channel
are so suppressed that they are unobservable \cite{Han:2006ip}. However
the rate of heavy neutrino pair-production via a $Z$ boson is independent
of this mixing, becoming the dominant production mechanism in the
small mixing regime. Moreover, if the mass difference between $\nu_{4}$
and $\nu_{\bar{4}}$ is at least 1 GeV, and the mixing so small that
the decay $\nu_{\bar{4}}\rightarrow\nu_{4}Z$ always dominates, the
above processes also lead to like-sign leptons in half of the events
\footnote{In the exact Dirac limit, $\nu_{\bar{4}}$ must decay to $W\ell$
and the different contributions to same sign di-lepton production
cancel, since the Dirac neutrino conserves lepton number. However,
as far as $\nu_{\bar{4}}$ always decays to $\nu_{4}Z$ there is no
interference amplitude, and same-sign di-lepton decays are unsuppressed.%
}. See ref.\cite{Rajaraman:2010ua} for a detailed study of the Tevatron
dataset potential to exclude (or discover) fourth generation neutrinos
with both, Dirac and Majorana masses, up to 150-175~GeV, depending
on the mixing. For the LHC, only the pure Majorana case has been studied
in ref.\cite{Rajaraman:2010wk}. According to them, the LHC at
$\sqrt{s}=$ 10 TeV with 5~${\rm fb}^{-1}$ could expect to set a 95\%
CL mass lower limit of $m_{N}>300\,\mathrm{GeV}$ or report 3$\sigma$
evidence for the $\nu_{4}$ if $m_{\nu_{4}}<225\,\mathrm{GeV}$. We
expect a similar sensitivity in our model, in the region $m_{4}-m_{\bar{4}}>1\,\mathrm{GeV}$
and small mixing ($\epsilon,\epsilon'\lesssim10^{-4}$) i.e., somehow
complementary to the one probed in LFV processes. See also \cite{CuhadarDonszelmann:2008jp}
for an evaluation of the LHC discovery potential for both Majorana
and Dirac type fourth family neutrinos in the process $pp\rightarrow Z/H\rightarrow\nu_{4}\bar{\nu}_{4}\rightarrow W\mu W\mu$.

Searches for fourth generation charged leptons at the LHC have been
studied in \cite{Carpenter:2010bs}, also in a general framework with
Dirac and Majorana neutrino masses, and assuming that the neutrino
$\nu_{4}$ is the lightest fourth generation lepton. For charged leptons
with masses under about 400 GeV, the dominant production channel is
charged lepton - neutrino, through the process $q\bar{q}'\rightarrow W^{\pm}\rightarrow\nu_{4}E^{\pm}$.
The neutrino $\nu_{4}$ can only decay to $\nu_{4}\rightarrow W\ell$,
and being Majorana it can decay equally to $W^{-}\ell^{+}$ and $W^{+}\ell^{-}$.
Therefore when a pair of fourth generation leptons are produced, we
expect the decay products to contain like-sign di-leptons half of
the time. The sensitivity study for this process in events with two
like-sign charged leptons and at least two associated jets shows that
with $\sqrt{s}=$ 7~TeV and 1~${\rm fb}^{-1}$ of data, the LHC can
exclude fourth generation charged lepton masses up to 250~GeV. It
would be interesting to study the parameter space in our model that
would lead to this type of signals.

In the above searches, it was assumed that the lightest neutrino decays
promptly. However, if the mixing of the lightest fourth generation
neutrino with the first three generation leptons is $\epsilon\lesssim10^{-7}$
its proper lifetime will be $\tau_{4}\gtrsim10^{-10}s$. The decay
length at the LHC is given by $d=\beta c\gamma\tau_{4}\sim 3\mathrm{cm}(\tau_{4}/10^{-10}s)\beta\gamma$,
thus for $\tau_{4}\gtrsim10^{-10}s$ the fourth neutrino will either
show displaced vertices in its decay or decay outside the detector,
if $d\gtrsim{\cal O}(m)$, which is a typical detector size. In our
scenario, such a tiny mixing is only compatible with large Majorana
masses, $m_{R}\sim$ 1 TeV (see figure~\ref{fig:Parameter-space}),
far from the pseudo-Dirac case. Searches for Majorana neutrinos stable
on collider times have been discussed in \cite{Carpenter:2010sm},
where it is proposed to use a quadri-lepton signal that follows from
the pair production and decay of heavy neutrinos $pp\rightarrow Z\rightarrow\nu_{\bar{4}}\nu_{\bar{4}}\rightarrow\nu_{4}\nu_{4}ZZ$,
when both $Z$'s decay leptonically. The final state is thus 4$\ell$
plus missing energy. For 30~${\rm fb}^{-1}$ of LHC data at 13~TeV,
$\nu_{4}$ masses can be tested in the range 100 to 180 GeV, and $\nu_{\bar{4}}$
masses from 150 to 250~GeV.

Finally, if the lightest fourth generation lepton is the charged one,
there is a striking signal which to our knowledge has not been studied
in the literature: lepton number violating like-sign fourth generation
lepton pair-production, through $q\bar{q}'\rightarrow W^{\pm}\rightarrow E^{\pm}\nu_{4,\bar{4}}\rightarrow E^{\pm}E^{\pm}W^{\mp}$
or via $W$ fusion, $q\bar{q}\rightarrow W^{\pm}W^{\pm}q'q'\rightarrow E^{\pm}E^{\pm}jj$
These processes are not suppressed by the small mixing with the first
three generations, so in principle they could be observable in our
scenario. Depending on the charged lepton lifetime, they will decay
promptly to same-sign light di-leptons, show displaced vertices or
leave an anomalous track of large ionization and/or low velocity.
A detailed phenomenological study would be very interesting, but it is beyond
the scope of this work.

\section{Summary and conclusions\label{conclusions}}

We have analysed a simple extension of the SM in which light neutrino
masses are linked to the presence of $n$ extra generations with both
left- and right-handed neutrinos. The Yukawa neutrino matrices are
rank $n$, so if we do not impose lepton number conservation and allow
for right-handed neutrino Majorana masses, at tree level there are
2$n$ massive Majorana neutrinos and three massless ones. In order
to obtain heavy neutrino masses above the experimental limits from
direct searches at LEP, the Dirac neutrino masses should be at the
electroweak scale, similar to those of their charged lepton partners,
and the right-handed neutrino Majorana masses can not be too high
(of order 10 TeV at most). 
The three remaining neutrinos get Majorana
masses at two loops, therefore this framework provides
a natural explanation
for the tiny masses of the known SM neutrinos. On the other hand, 
it should be kept in mind that the two-loop contribution to
the neutrino mass matrix is always present in this type of SM
extensions, therefore the experimental upper limit on the absolute
light neutrino mass scale leads to a relevant constraint which
has to be taken into account.

We have shown that the minimal extension with a fourth generation
can not fit simultaneously the ratio of the solar and atmospheric
neutrino mass scales, $\Delta m_{sol}^{2}/\Delta m_{atm}^{2}$, the
lower bound on the heavy neutrino mass from LEP and the limits on
the mixing between the fourth generation and the first three from
low energy universality tests. Then, there are two possibilities:
either enlarge the Higgs sector \cite{Grimus:1999wm} or consider
a five generation extension. In this work we have analyzed the second
one, while the first will be studied elsewhere \cite{nosotros2doublets}.
Notice that five generations are still allowed by the combination
of collider searches for its direct production, indirect effects in
Higgs boson production at Tevatron and LHC, and precision electroweak
observables \cite{Novikov:2009kc}, provided the Higgs mass is roughly
$m_{H}>$~300~GeV. However they will be either discovered or fully
excluded at LHC, making our proposal falsifiable in the very near
future.

Given the large number of free parameters in a five generation framework
(10 neutrino Yukawa couplings, 2 charged lepton masses and 2 right-handed
neutrino Majorana masses), we have considered a very simple working
example assuming that i) the linear combinations of left-handed neutrinos
that get Dirac masses at tree level are orthogonal to each other,
ii) each extra generation left-handed neutrino couples only to one
of the two right-handed SM singlet states and iii) each extra generation
charged lepton couples only to one linear combination of the (tree-level)
massless neutrinos. Then, we are left with 2 neutrino Yukawa couplings
$y_{E},y{}_{F}$, 2 charged lepton masses $m_{E}$, $m_{F}$, 2 right-handed
neutrino Majorana masses $m_{aR}$,$a=4,5$, which characterize the
amount of total lepton number breaking, and two small parameters $\epsilon,\epsilon'$
which determine the mixing among the first three generations and the
new ones. Moreover, at leading order the two-loop neutrino masses
$m_{2}$, $m_{3}$ depend only on $\epsilon,m_{4R},y_{E},m_{E}$ and
$\epsilon',m_{5R},y{}_{F},m_{F}$, respectively (see eq.~\eqref{eq:m2m3-estimate}).
Even in this over-simplified case we are able to accommodate all current
data, including the observed pattern of neutrino masses and mixings
both for normal and inverted hierarchy spectrum. A definite prediction
of the model (independent of the above simplifying assumptions) is
that the three light neutrinos can not be degenerate.

We have explored the parameter space regions able to generate the
correct scale of neutrino masses, $\sim0.05$~eV. We find that for
typical values of $m_{E},m_{4D}$ ($m_{F},m_{5D}$) at the electroweak
scale, we need $\epsilon^{2}m_{4R}$ ($\epsilon'^{2}m_{5R})\lesssim$
1 keV to obtain the atmospheric mass scale (see figure~\eqref{fig:Parameter-space}).

We have also studied the current bounds on the mixing parameters $\epsilon$
and $\epsilon'$ from the non-observation of LFV rare decays $\ell_{\alpha}\rightarrow\ell_{\beta}\gamma$,
as well as from universality tests. All of them are independent of
the Majorana masses $m_{i R}$, since they conserve total lepton number.
Depending on the light neutrino mass spectrum (normal or inverted),
the strongest bounds come from $\mu\rightarrow e\gamma$ and from
universality tests in $\pi$ decays. Combining the information from
both processes we can set independent limits on $\epsilon$ and $\epsilon^{\prime}$
which being quite conservative are of the order of the few percent,
$\epsilon\lesssim0.03$ and $\epsilon^{\prime}\lesssim0.04$. 

Finally, we have analysed the phenomenological prospects of the model.
With respect to LFV signals, future MEG data will improve the limits
on the $\epsilon$'s by a factor of about 3 while, if expectations
from $\mu$--$e$ conversion are attained the limits on the $\epsilon$'s
will be pushed to $10^{-3}$ . This region of observable LFV effects
corresponds to the pseudo-Dirac limit, $m_{aR}\lesssim$ 1 GeV, i.e.,
two pairs of strongly degenerate heavy neutrinos. In this regime,
they can only be discovered at LHC using pure Dirac neutrino signatures,
which are more difficult to disentangle from the background.

On the other hand, we find that in the complementary region of very
small mixing $\epsilon,\epsilon'\ll10^{-3}$, $m_{aR}\gtrsim$ 1 GeV,
the lighter Majorana neutrinos $\nu_{4},\nu_{\bar{4}}$ will lead
to observable same-sign di-lepton signatures at LHC. A detailed study
is missing, but previous results seem to indicate that a lower bound
on $m_{4}$ of order 300 GeV could be set with 5~fb$^{-1}$ of LHC
data at 10 TeV \cite{Rajaraman:2010wk}.

\section{Acknowledgments}

We thank M. Hirsch and E. Fernández-Martínez for useful discussions.
This work has been partially supported by the Spanish MICINN under
grants FPA-2007-60323, FPA-2008-03373, Consolider-Ingenio PAU (CSD2007-00060)
and CPAN (CSD2007- 00042), by Generalitat Valenciana grants PROMETEO/2009/116
and PROMETEO/2009/128 and by the European Union within the Marie Curie
Research \& Training Networks, MRTN-CT-2006-035482 (FLAVIAnet). A.A.
and J.H.-G. are supported by the MICINN under the FPU program.

\appendix

\section{The neutrino mass two-loop integral\label{sec:appendixA}}

The relevant integral is \[
J\equiv J(m_{4},m_{\bar{4}},m_{\alpha},m_{\beta},m_{W})=\]
\begin{equation}
=\int_{pq}\cfrac{p\cdot q}{((p+q)^{2}-m_{4}^{2})((p+q)^{2}-m_{\bar{4}}^{2})(p^{2}-m_{\alpha}^{2})(q^{2}-m_{\beta}^{2})(p^{2}-m_{W}^{2})(q^{2}-m_{W}^{2})}\,,\label{eq:integral-pq}\end{equation}
where \[
\int_{pq}=\int\int\frac{d^{4}p}{(2\pi)^{4}}\frac{d^{4}q}{(2\pi)^{4}}\,.\]
We combine propagators with the same momentum by using \[
\frac{1}{(p^{2}-m_{\text{\ensuremath{\alpha}}}^{2})(p^{2}-m_{W}^{2})}=\frac{1}{(m_{\alpha}^{2}-m_{W}^{2})}\int_{m_{W}^{2}}^{m_{\alpha}^{2}}\frac{dt_{1}}{(p^{2}-t_{1})^{2}}\,,\]
then\[
J=\int_{t}\int_{pq}\frac{(pq)}{(p^{2}-t_{1})^{2}(q^{2}-t_{2})^{2}((p+q)^{2}-t_{3})^{2}}\,,\]
where \[
\int_{t}=\frac{1}{(m_{\alpha}^{2}-m_{W}^{2})}\frac{1}{(m_{\beta}^{2}-m_{W}^{2})}\frac{1}{(m_{4}^{2}-m_{\bar{4}}^{2})}\int_{m_{W}^{2}}^{m_{\alpha}^{2}}dt_{1}\int_{m_{W}^{2}}^{m_{\beta}^{2}}dt_{2}\int_{m_{4}^{2}}^{m_{\bar{4}}^{2}}dt_{3}\,.\]
Now we use the standard Feynman parametrization to combine the last
two propagators and perform the integral in $q$, which leads to\[
J=-\frac{i}{(4\pi)^{2}}\int_{t}\int_{0}^{1}\frac{dx}{1-x}\int_{p}\frac{p^{2}}{(p^{2}-t_{1})^{2}(p^{2}-t_{3}/(1-x)-t_{2}/x)^{2}}\,.\]
The integral in $p$ can be reduced by using an additional Feynman
parameter and the final result can be written as\begin{equation}
J=-\frac{2}{(4\pi)^{4}}\int_{0}^{1}dx\int_{0}^{1}dy\int_{t}\frac{y(1-y)x}{t_{3}xy+t_{2}(1-x)y+t_{1}x(1-x)(1-y)}\,.\label{eq:IntegralFeynman}\end{equation}
The integrals in $t_{1},t_{2},t_{3}$ can be done analytically and
reduced to logarithms. The expressions obtained are complicated but
can be used to feed the final numerical integration in $x$ and $y$
which converges smoothly for most of the parameters. The expression
in \eqref{eq:IntegralFeynman} is also very useful to obtain different
approximations for small masses as compared with the largest mass
in the integral. For that purpose one can use\[
\lim_{a\rightarrow0}\lim_{b\rightarrow a}\frac{1}{b-a}\int_{a}^{b}dtf(t)=\lim_{a\rightarrow0}f(a)=f(0)\,.\]
Thus, for instance if $m_{\bar{4}},m_{4}\gg m_{\alpha},m_{\beta},m_{W}\sim0$
we can take $t_{1},t_{2}\rightarrow0$ in the integrand and perform
trivially the remaining integrals,\begin{equation}
J=-\frac{1}{(4\pi)^{4}}\frac{1}{m_{\bar{4}}^{2}-m_{4}^{2}}\ln\frac{m_{\bar{4}}^{2}}{m_{4}^{2}},\label{eq:J1}\end{equation}
in agreement with the result in \cite{Babu:1988wk}. 

If $m_{\beta},m_{\bar{4}},m_{4}\gg m_{\alpha},m_{W}\sim0$ the integral
can also be computed (take $t_{1}\rightarrow0$ in the integrand and
perform the rest of the integrals). The result can be written in terms
of the dilogarithm function $\mathrm{Li}_{2}(x)$ and it is rather
compact, \[
J=-\frac{1}{(4\pi)^{4}m_{\beta}^{2}}\left(\frac{\pi^{2}}{6}-\frac{m_{\bar{4}}^{2}}{m_{\bar{4}}^{2}-m_{4}^{2}}\left(\mathrm{Li}_{2}\left(1-\frac{m_{\beta}^{2}}{m_{\bar{4}}^{2}}\right)-\frac{m_{4}^{2}}{m_{\bar{4}}^{2}}\mathrm{Li}_{2}\left(1-\frac{m_{\beta}^{2}}{m_{4}^{2}}\right)\right)\right)\,.\]
When $m_{\beta}\rightarrow0$ it reduces, as it should, to \eqref{eq:J1}
.

We are especially interested in the case $m_{\alpha}=m_{\beta}\equiv m_{E}$
with $m_{E}>m_{W}$, but $m_{4},m_{\bar{4}}$ could be larger or smaller
than $m_{E}$ (and even smaller than $m_{W}$ since we only know that
$m_{\bar{4}}\ge m_{4}>63\,\mathrm{GeV}$). Some asymptotic expressions
can be obtained when there are large hierarchies in masses\[
J\approx-\frac{1}{(4\pi)^{4}}\frac{1}{m_{X}^{2}}\ln\frac{m_{X}^{2}}{m_{Y}^{2}}\;,\]
where $m_{X}$ is the heaviest of $m_{E},m_{\bar{4}},m_{4},m_{W}$
and $m_{Y}$ the next to the heaviest of these masses. This expression
can be used to perform analytical estimates but, since in the allowed
range of masses the hierarchies cannot be huge, we do expect large
corrections to these estimates. Fortunately, as commented above, the
exact value of $J$ for all values of the masses can be obtained numerically
rather easily using \eqref{eq:IntegralFeynman}. For fast estimates
one can use\[
J\approx\frac{1}{(4\pi)^{4}}\frac{1}{m_{\text{\ensuremath{\bar{4}}}}^{2}-m_{4}^{2}-m_{E}^{2}}\ln\left(\frac{m_{4}^{2}+m_{E}^{2}}{m_{\text{\ensuremath{\bar{4}}}}^{2}}\right)\;,\qquad m_{E},m_{\bar{4}},m_{4}\gg m_{W},\]
which interpolates smoothly the different asymptotic expressions and
reproduces the complete result with an error less than $50\%$ in
the worse case. 

\providecommand{\href}[2]{#2}\begingroup\raggedright\endgroup

%\bibliographystyle{JHEP}
%\bibliography{4gens}

\end{document}